\documentclass[conference]{IEEEtran}
\usepackage{amsmath}
\usepackage{amsthm}
\usepackage{algorithm}
\usepackage{algorithmic}
\usepackage{subfigure}
\usepackage{graphicx}
\usepackage{epstopdf} 
\usepackage{float}  
\usepackage{hyperref}

\graphicspath{{figures_svd/}}

\usepackage{verbatim} 

\newcommand{\n}{|\!|}

\renewcommand\IEEEkeywordsname{Index Terms}

\ifCLASSINFOpdf
\else
\fi
\begin{document}
%
\title{SVD-based Visualisation and Approximation for Time Series Data in Smart Energy Systems
\thanks{This work was supported in part by the Dutch STW under project grant \emph{Smart Energy Management and Services in Buildings and Grids (SES-BE)}.}
}

\author{\IEEEauthorblockN{Abdolrahman Khoshrou}
\IEEEauthorblockA{Centrum Wiskunde \& Informatica\\
Science Park 123, 1098 XG\\ Amsterdam, The Netherlands\\
Email: a.khoshrou@cwi.nl
}
\and
\IEEEauthorblockN{Andr\'e B. Dorsman}
\IEEEauthorblockA{Vrije Universiteit Amsterdam, \\De Boelelaan 1105, 1081 HV \\Amsterdam, The Netherlands\\
Email: a.b.dorsman@vu.nl}
\and
\IEEEauthorblockN{Eric J. Pauwels}
\IEEEauthorblockA{Centrum Wiskunde \& Informatica\\
Science Park 123, 1098 XG\\ Amsterdam, The Netherlands\\
Email: eric.pauwels@cwi.nl}}
%
\IEEEpubid{\makebox{{978-1-5386-1953-7/17/\$31.00~\copyright~2017 IEEE}}}
\maketitle
\pagestyle{empty} 
\begin{abstract} 
\boldmath 
Many time series in smart energy systems exhibit two different timescales.  On the one hand there 
are patterns linked to daily human activities. On the other hand, there are relatively slow trends linked 
to seasonal variations.  In this paper we interpret these time series as matrices,  
to be visualized as images.  This approach has two advantages: 
First of all, interpreting such time series as images enables one to visually integrate across the image and makes it therefore easier to spot subtle or faint features.
Second, the matrix interpretation also grants elucidation of the underlying structure using well-established matrix decomposition methods.
We will illustrate both these aspects for data obtained from the German day-ahead market. 
\end{abstract}
%
\IEEEpeerreviewmaketitle
\section{Introduction}
In smart energy systems, time series often show two distinct time scales. 
On the one hand, the data exhibit strong  diurnal patterns reflecting the daily (or weekly) rhythms of human 
activity.  
On the other hand, these relatively fast diurnal patterns are superimposed on slower seasonal variations that have
a significant impact on the overall evolution of the data. 
To improve the visualization and make it easier to spot correlations between variables, we propose to analyze these time series as matrices (to be visualized as images) where rows represent hour slots, whereas columns correspond to days.  
This approach has two advantages.  First,  visualizing such time series as images allows one to 
visually integrate across the image and makes it therefore easier to spot subtle or faint features. 
Second, one can draw on well-established matrix decomposition methods to elucidate underlying structure. 
In this paper we will discuss both these aspects in the context of data from the day-ahead market.  
\section{Data} \label{sec:data}
\begin{figure}[!h]
    \centering
    \includegraphics[width=7.5cm,height=3.3cm]{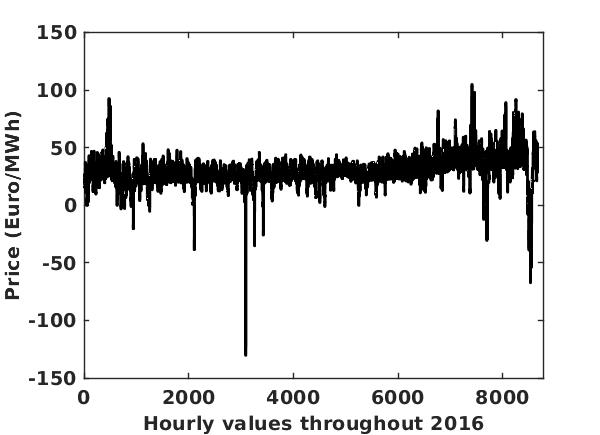}
    \includegraphics[width=7.5cm,height=3.3cm]{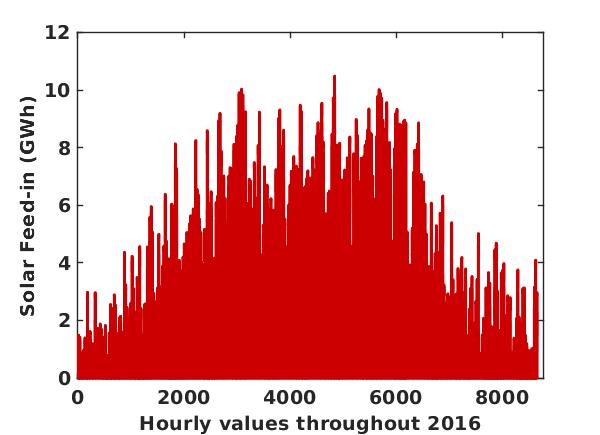}
    \includegraphics[width=7.5cm,height=3.3cm]{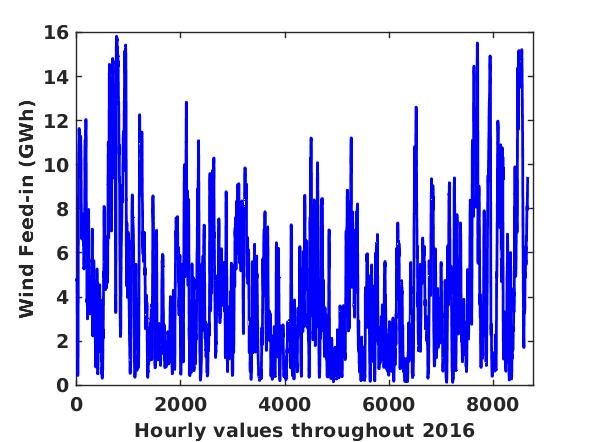}
    \includegraphics[width=7.5cm,height=3.3cm]{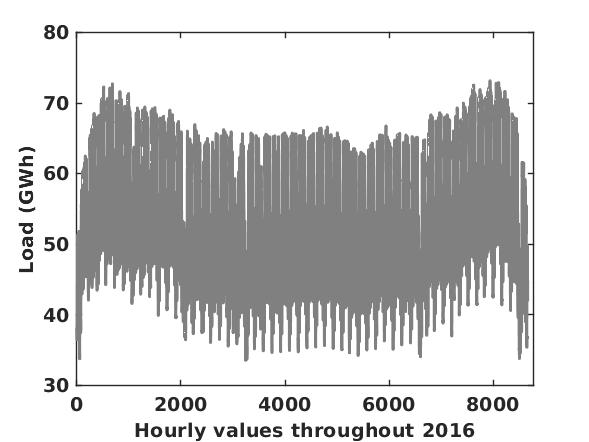}
    \includegraphics[width=7.5cm,height=3.3cm]{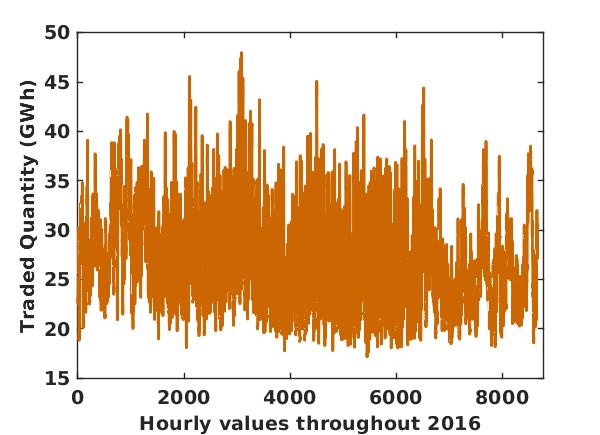}
    \caption{An overview of the German day-ahead market in 2016; each data point represents one hour slot.  From top to bottom, the price, solar, wind, load, and the traded quantity.}
    \label{fig:GE_day_ahead_timeseries}
\end{figure}

The day-ahead market is an exchange for short-term electricity contracts where the tradings are driven by its participants~{\cite{epex}}.
Fig.~\ref{fig:GE_day_ahead_timeseries} illustrates various collected sets of data for the German day-ahead market in 2016.  
Day-ahead price and the traded quantity data were collected from~{\cite{epexPrice}}. 
We obtained the day-ahead solar and wind feed-in energy data from~{\cite{Tennet}}. 
ENTSO-E, the European Network of Transmission System Operators~{\cite{entsoe}}, was the platform for downloading the day-ahead load forecast.

Using the above-mentioned data, we will explore the influence of the daily fluctuations of the predicted supply of renewable energy sources (RES), viz. wind and solar feed-in, on the realized  electricity price dynamics. 
\IEEEpubidadjcol
%
\section{Background and Literature Review}  \label{sec:background}
Recently, the impact of variable generation on the electricity market has attracted a lot of attention.  
Denny et al.~{\cite{denny2010impact} } 
explore how increased interconnection between Great Britain and Ireland would facilitate the integration of the wind farms into the power system. 
Simulation results in this work imply that large increases in the interconnection capacities bring about a decrease in average price and its volatility in Ireland.
Furthermore, the growing contribution of intermittent energy sources enforces the transmission grid extensions and expanding the cross-border interconnections capacities to ensure the grid stability.
K. Schaber, F. Steinke, and T. Hamacher in~{\cite{schaber2012transmission}} examine the viability of this approach and its effects, based on the projected wind and solar data until 2020. 
They conclude that grid expansion is, indeed, helpful in coping with externalities which come with the deployment of renewable energies.  
The  positive outcomes of the substantial deployment of photovoltaic (PV) installations in Germany and Italy, and in particular,  their role in daytime peak price drop have been discussed by
K. Barnham, K. Knorr, and M. Mazzer in~{\cite{barnham2013benefits}}. 
This work also reports the benefits of the complementary nature of wind and PV resources in the UK.
Continuing further with studying the influence of renewable energy sources (RES) in Germany, a preliminary study on the German day-ahead market has been carried out in~{\cite{adaduldah2014influence}}. 
In the reported work, N. Adaduldah, A. Dorsman, GJ. Franx, and P. Pottuijt have taken into account the priority that the German policies assign to renewables over fossil fuels in case of adequate supply.  
The authors reported the existence of convincing evidence for the impact of RES on the recent emergence of negative prices on the German day-ahead market.

Inspired by the work in~{\cite{hirth2013market}}, the goal of the present work is to determine the influence of the variability of the wind and solar feed-in on the price variability in the day-ahead market. To this end we focus on the intra-day dynamics of price as characterized by its second derivative, as this peaks for sharp trend reversals. 
%
%
%
\section{Methods: Analysing Time Series as Images} \label{sec:image}
\subsection{Motivation}
\begin{figure}[!]
    \centering
    \includegraphics[width=7.5cm,height=3.3cm]{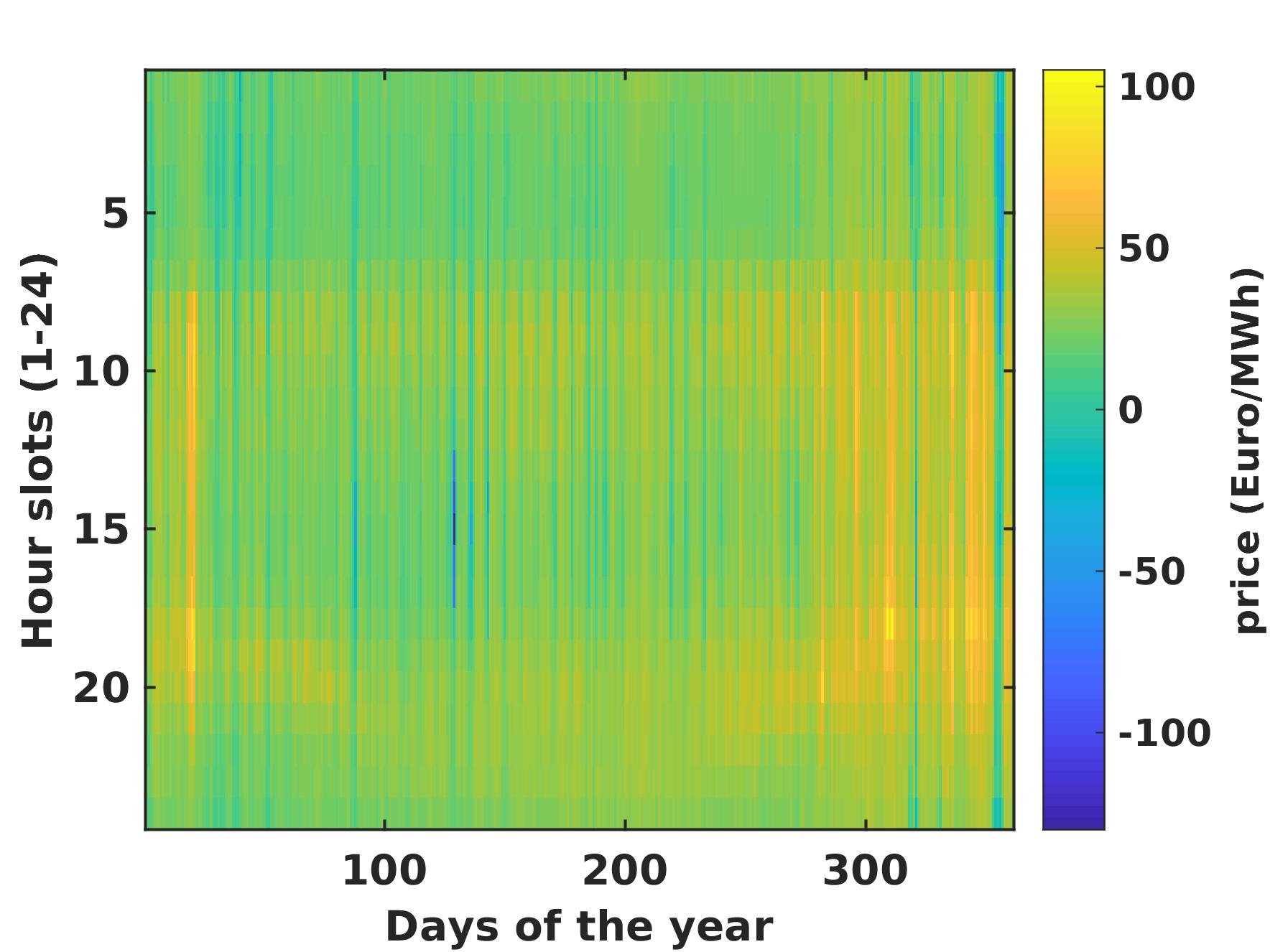}
    \includegraphics[width=7.5cm,height=3.3cm]{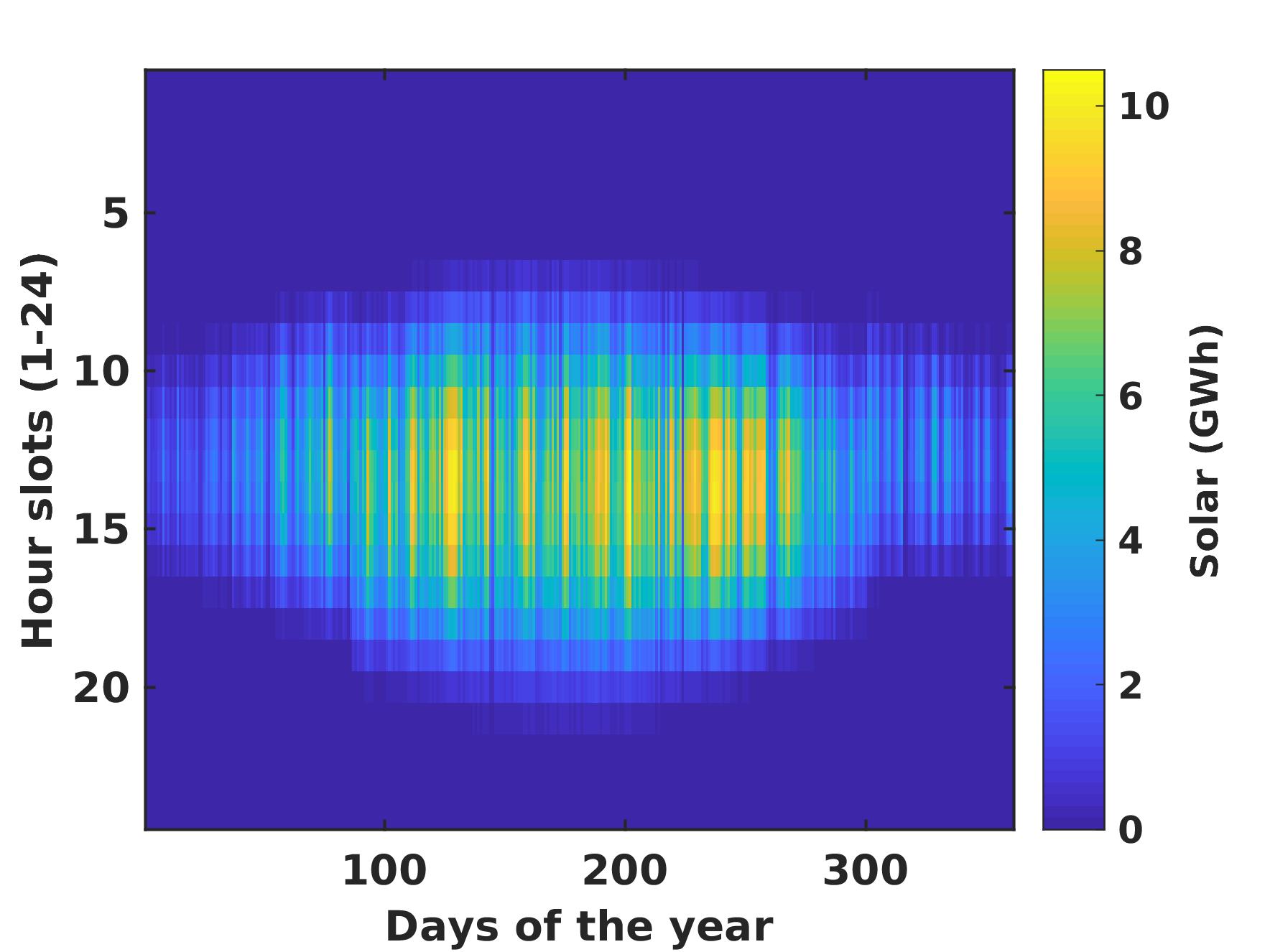}
    \includegraphics[width=7.5cm,height=3.3cm]{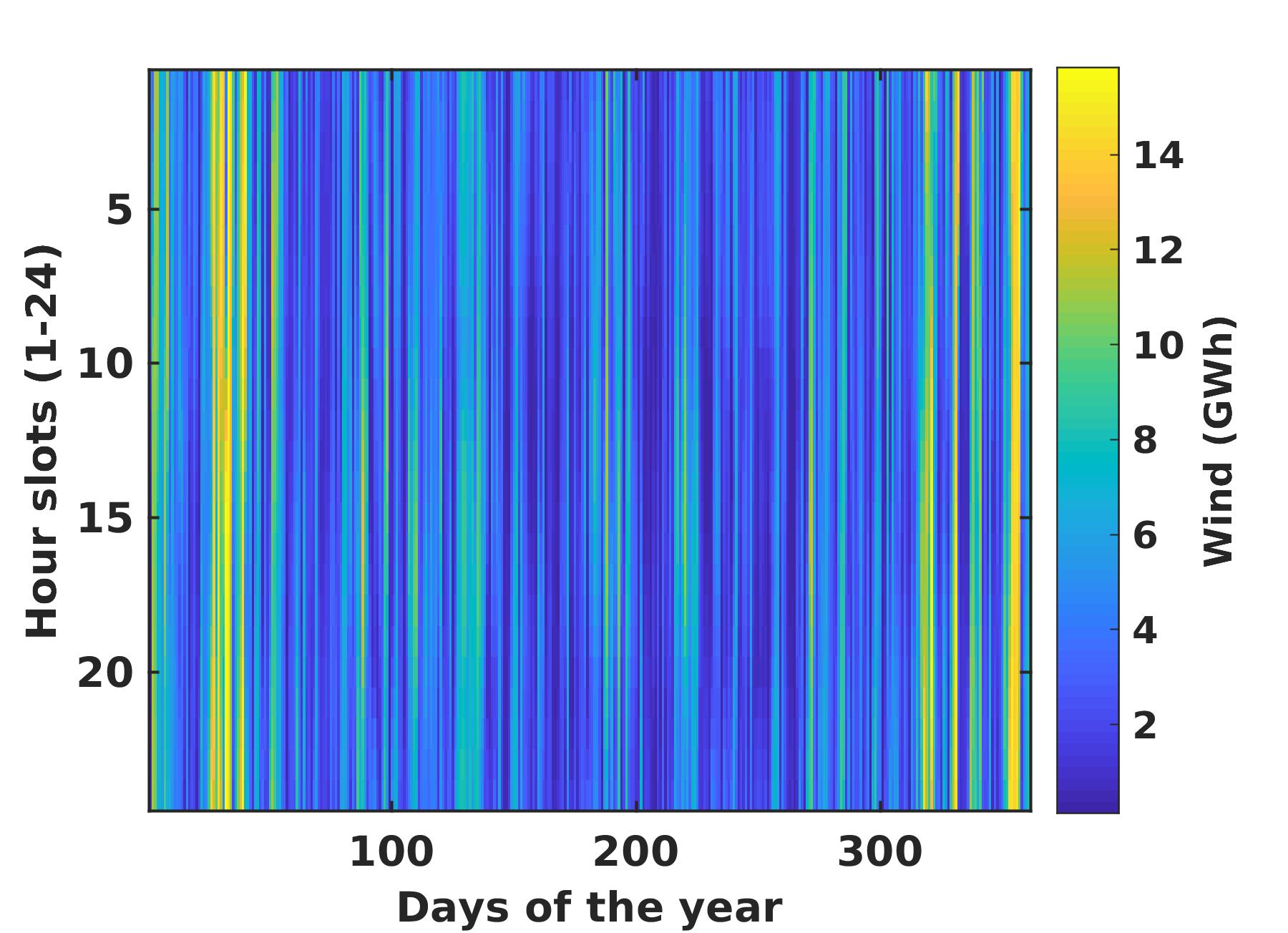}
    \includegraphics[width=7.5cm,height=3.3cm]{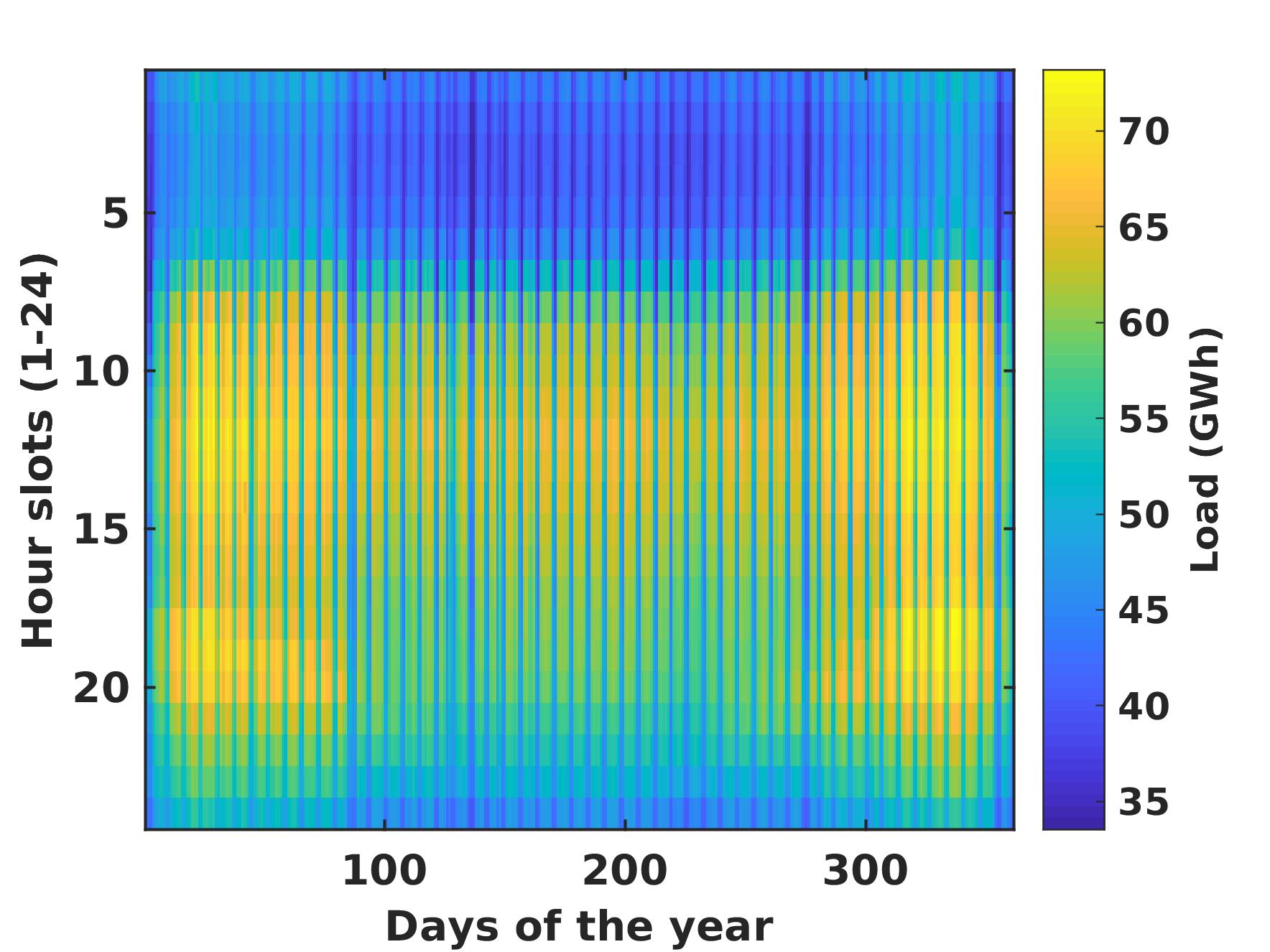}
    \includegraphics[width=7.5cm,height=3.3cm]{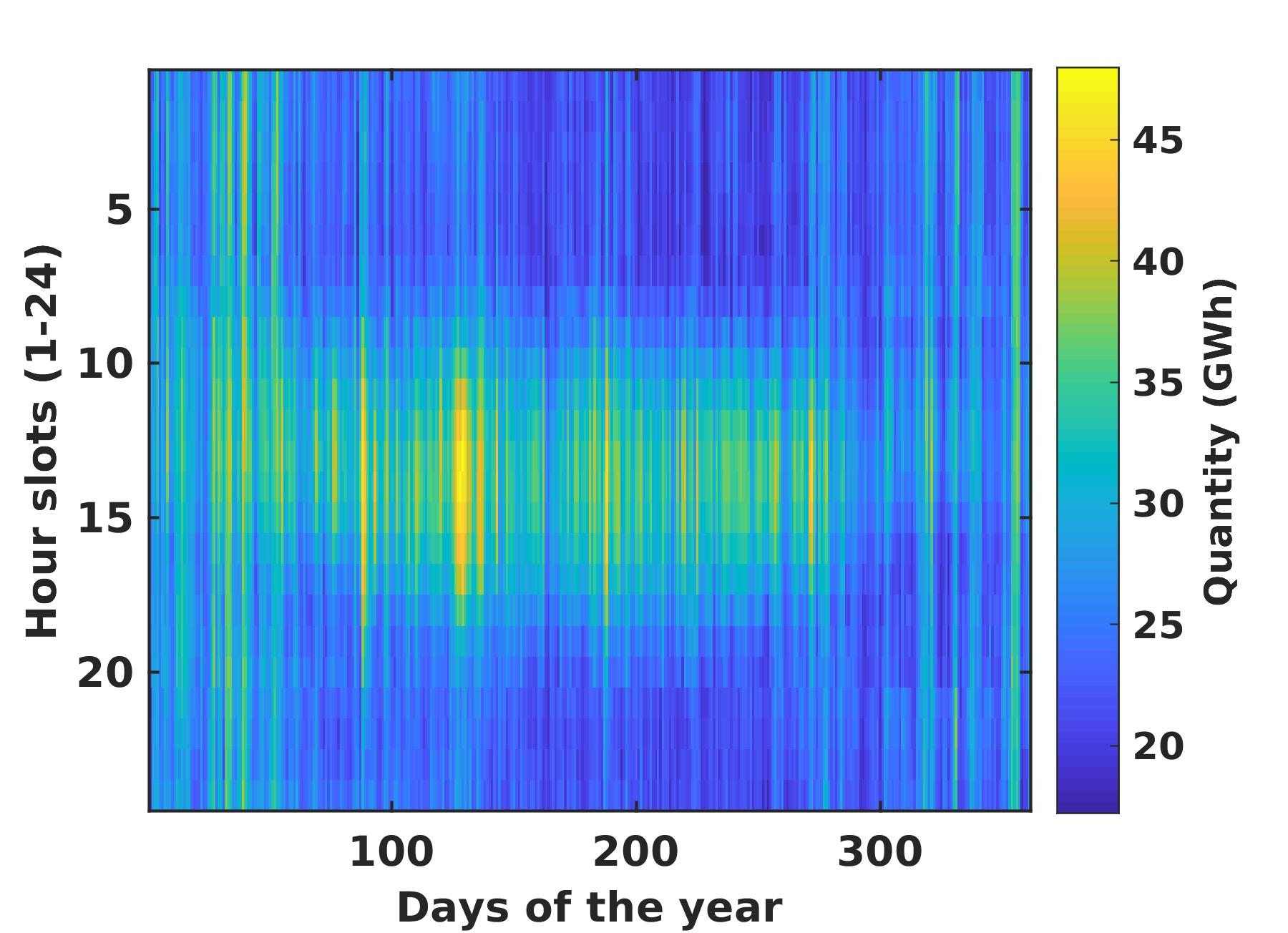}
    \caption{A more informative representation of the German day-ahead market by reformatting the timeseries in Fig.~\ref{fig:GE_day_ahead_timeseries} into matrices of size $24 \times 366$. 
    Each image (or matrix) column represents a single day of 24 hour-values.
    From top to bottom, the price, solar, wind, load, and the traded quantity.}
    \label{fig:GE_day_ahead_images}
\end{figure}

An alternative way of visualizing time series with diurnal patterns is as matrices~\cite{kanjilal1994singular}. Fig.~\ref{fig:GE_day_ahead_images} shows one year's worth of data represented  as a $24\times 366$ matrix/image (recall that 2016 is a leap year). 
%
%
An important distinction of this new way of representation is that it allows us to visually integrate patterns across longer time spans, which results in higher discriminatory power.  
For instance, even a cursory glance at the data image for the traded quantity 
(Fig.~\ref{fig:GE_day_ahead_images}, bottom) highlights the fact that there is a significant correlation with solar (the eye-like horizontal shape seen in the 2nd panel from the top) as well as wind (the vertical stripes in the 3rd panel from the top). 
\subsection{Finding peaks and valleys}
\begin{figure}[!h]
    \centering
    \includegraphics[width=7cm,height=4cm]{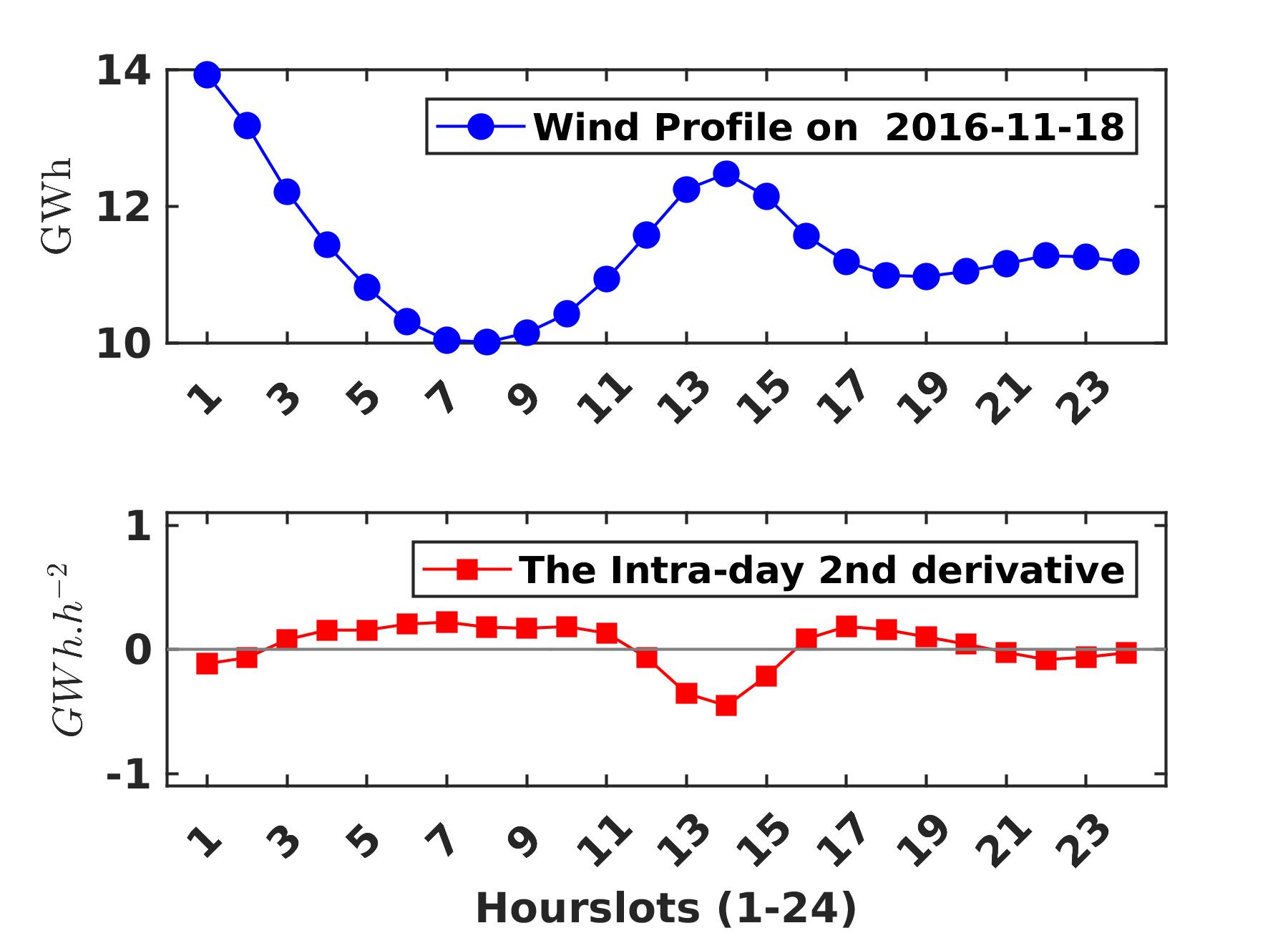}
    \caption{Wind profile and its corresponding intra-day 2nd derivative on Nov. 18, 2016. 
    Every sag in the lower profile corresponds to a swell in upper and vice versa.}
    \label{fig:main_20170302_svd_wind_2nd_derivative}
\end{figure}

As mentioned before, the main scope of this work is to explore how the inherent variability of the supply by RES can affect the intra-day variability of the market.
To address this problem, we have applied an additional transformation on the daily profiles of the quantities of interest (viz. price, load, traded quantity, solar and wind feed-in). 
More precisely, any of the above quantities (generically denoted by $f$) can be considered as a function of two variables:  
\begin{itemize}
    \item time of day, hour slots $1 \leq h \leq 24$
    \item day of year, $1\leq d \leq 366$ (2016 is a leap year!)
\end{itemize}
Hence for such a function $f(h,d)$, we can investigate the corresponding 2nd derivative with respect to the hour (intra-day); 
\begin{equation}
    f_{hh}\equiv \frac{\partial^2 f}{\partial h^2} \approx \frac{f(h+1)-2f(h)+f(h-1)}{h^2}
    \label{2nd_derivative}
\end{equation}
%
Extreme values of this 2nd derivative capture peaks (i.e. local maxima for which $f_{hh} < 0$ and extreme) or valleys (i.e. local minima for which $  f_{hh} > 0$ and extreme). 
An example is illustrated in Fig.~\ref{fig:main_20170302_svd_wind_2nd_derivative} which shows the wind feed-in profile (top) and its corresponding intra-day 2nd derivative profile (bottom) on Nov. 18, 2016.
We contend that comparing the evolution of the 2nd derivative profiles (intra-day wise) can highlight the impact of the wind and solar energy feed-in on the price and also the traded quantity in 2016.

\begin{figure}[]
    \centering 
      \includegraphics[width=7.5cm,height=3.3cm]{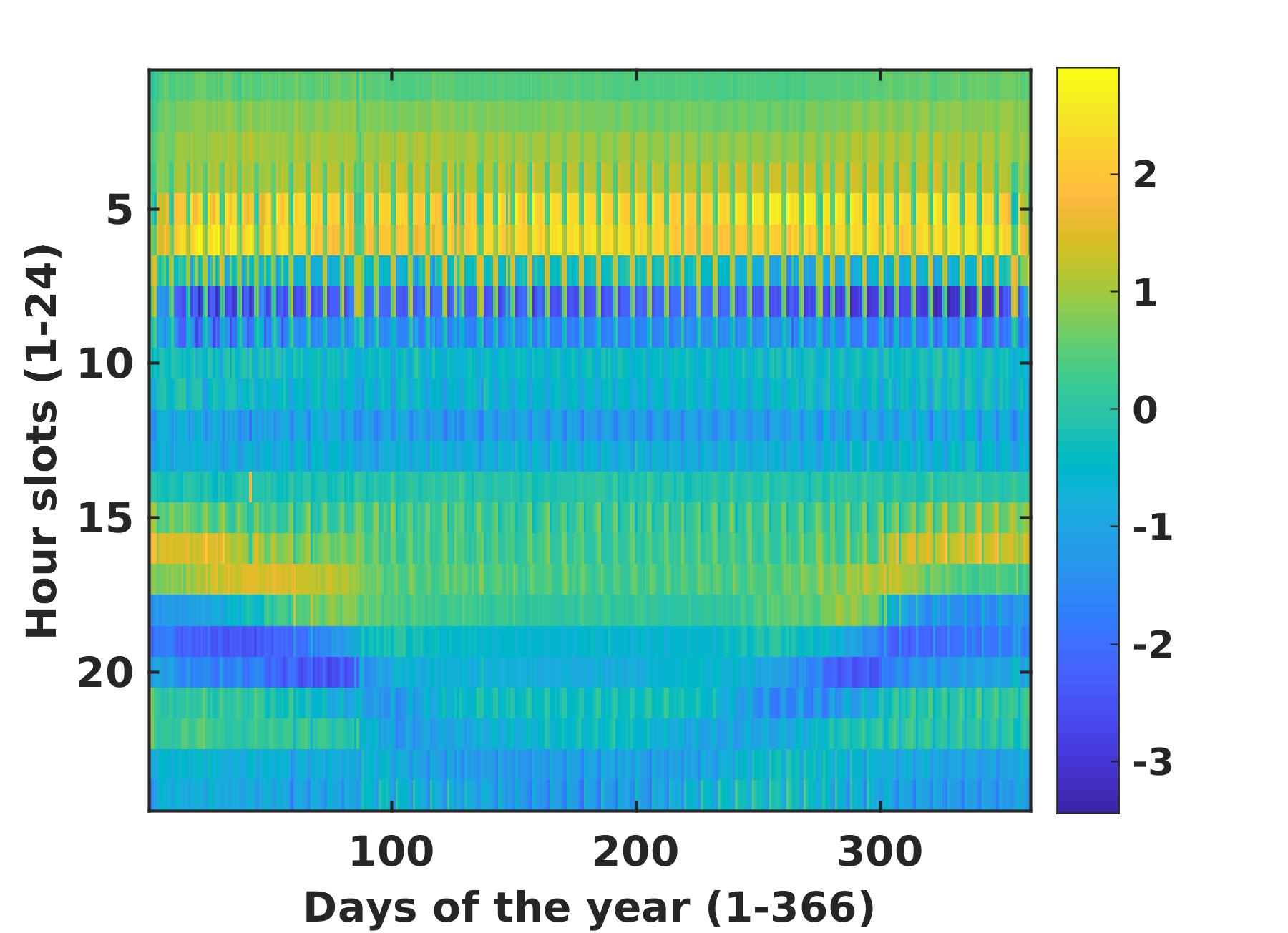}
      \includegraphics[width=7.5cm,height=3.3cm]{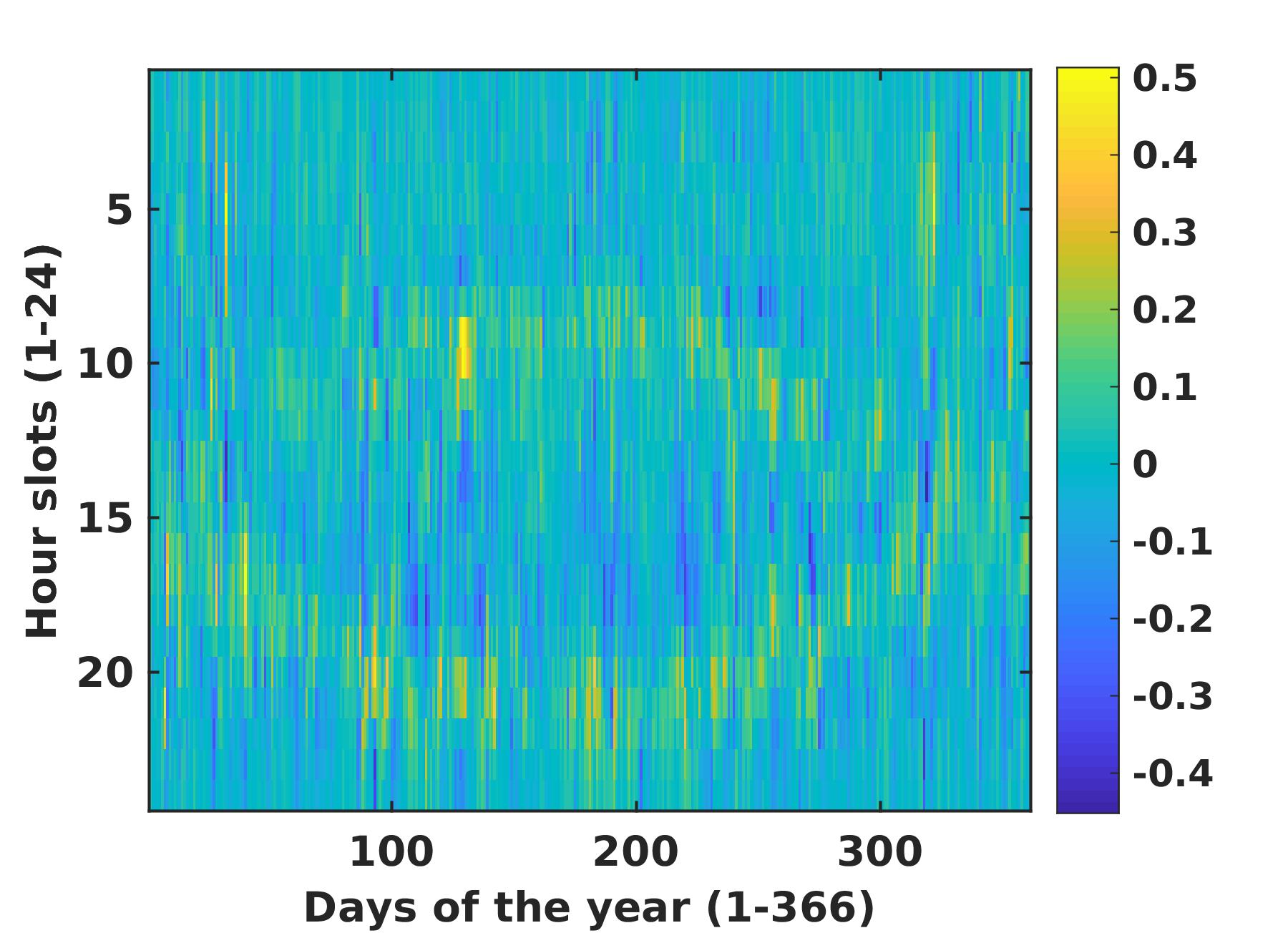}
    \caption{The evolution of appropriately transformed hourly values (2nd derivatives, see text for more details) for solar (top) and wind (bottom) feed-in.  
    Top:  Solar feed-in (intra-day 2nd derivative).
    The colour coding (in yellow) clearly highlights the change in sunrise and sunset times, 
    creating an overall ``eye-like'' structure. 
    Bottom: Wind feed-in (intra-day 2nd derivative).  As expected, wind feed-in is 
    much more erratic. Interestingly enough, this figure shows a vague but undeniable outline of an 
    eye-like contour which mirrors the intra-day wind activities. 
    }
    \label{fig:main_130217_wind_solar}
\end{figure}

Applying this intra-day 2nd derivative operator to all five quantities of interest and 
re-visualising the resulting time series as images highlights some interesting features of the data. 
As an illustration, Fig.~\ref{fig:main_130217_wind_solar}  shows the 2nd derivative for 
both solar (top) and wind feed-in (bottom). 
In the top panel the gradual shift of sunrise and sunset over the seasons is clearly visible. Moreover, close inspection of this figure also reveals the dates of the switch to daylight saving summer time (days 87 and 304). 
The image for the wind feed-in is also interesting. As expected, wind values are more erratic and less seasonally determined. That being said, there is an undeniable ``eye-like'' shape that faintly mirrors the intra-day wind activities (see, e.g.,~{\cite{knmi}}).
%
\subsection{Using SVD to highlight structure}
\begin{figure}[]
    \centering 
    \includegraphics[width=7.5cm,height=3.3cm]{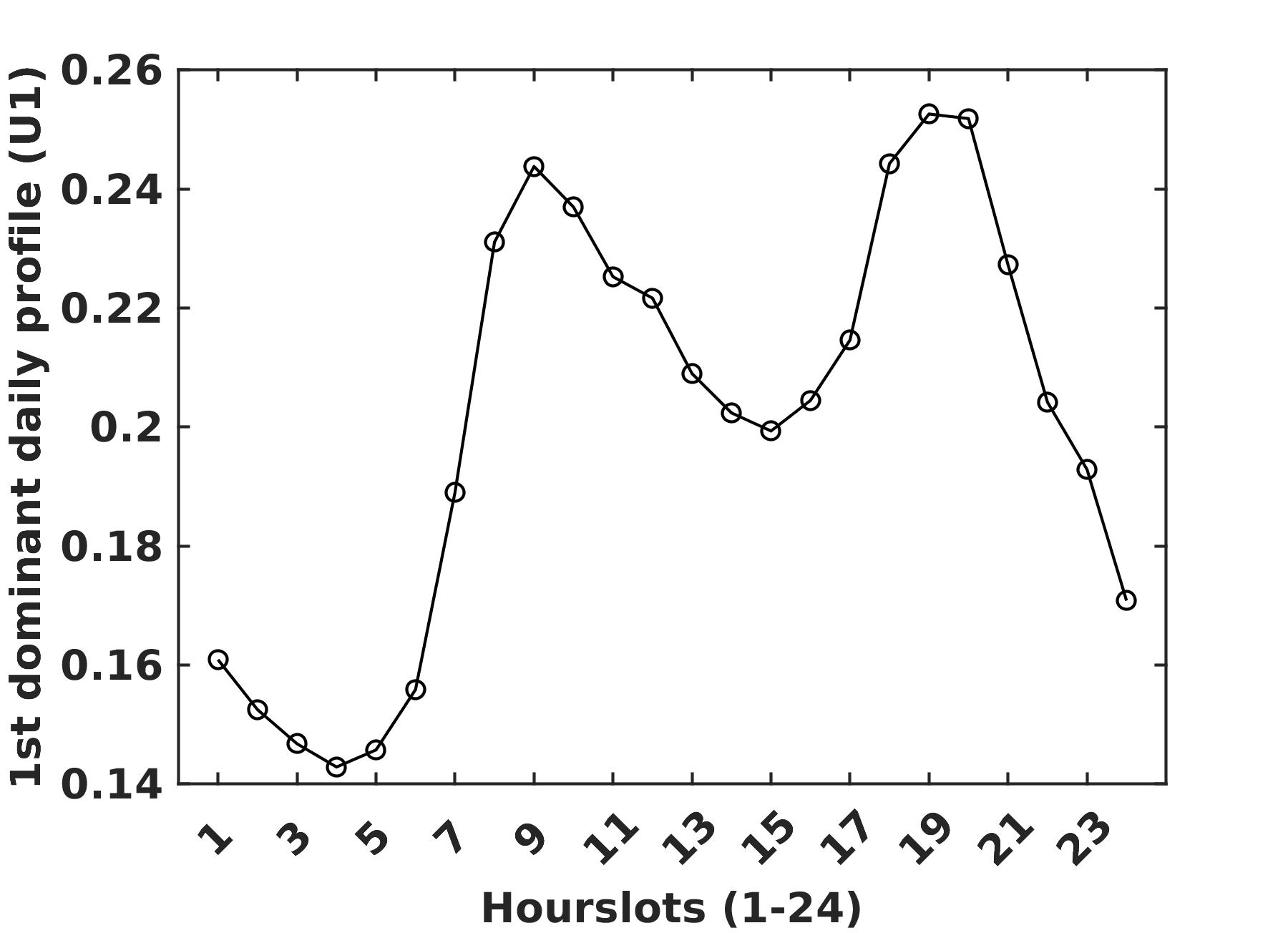}
    \includegraphics[width=7.5cm,height=3.3cm]{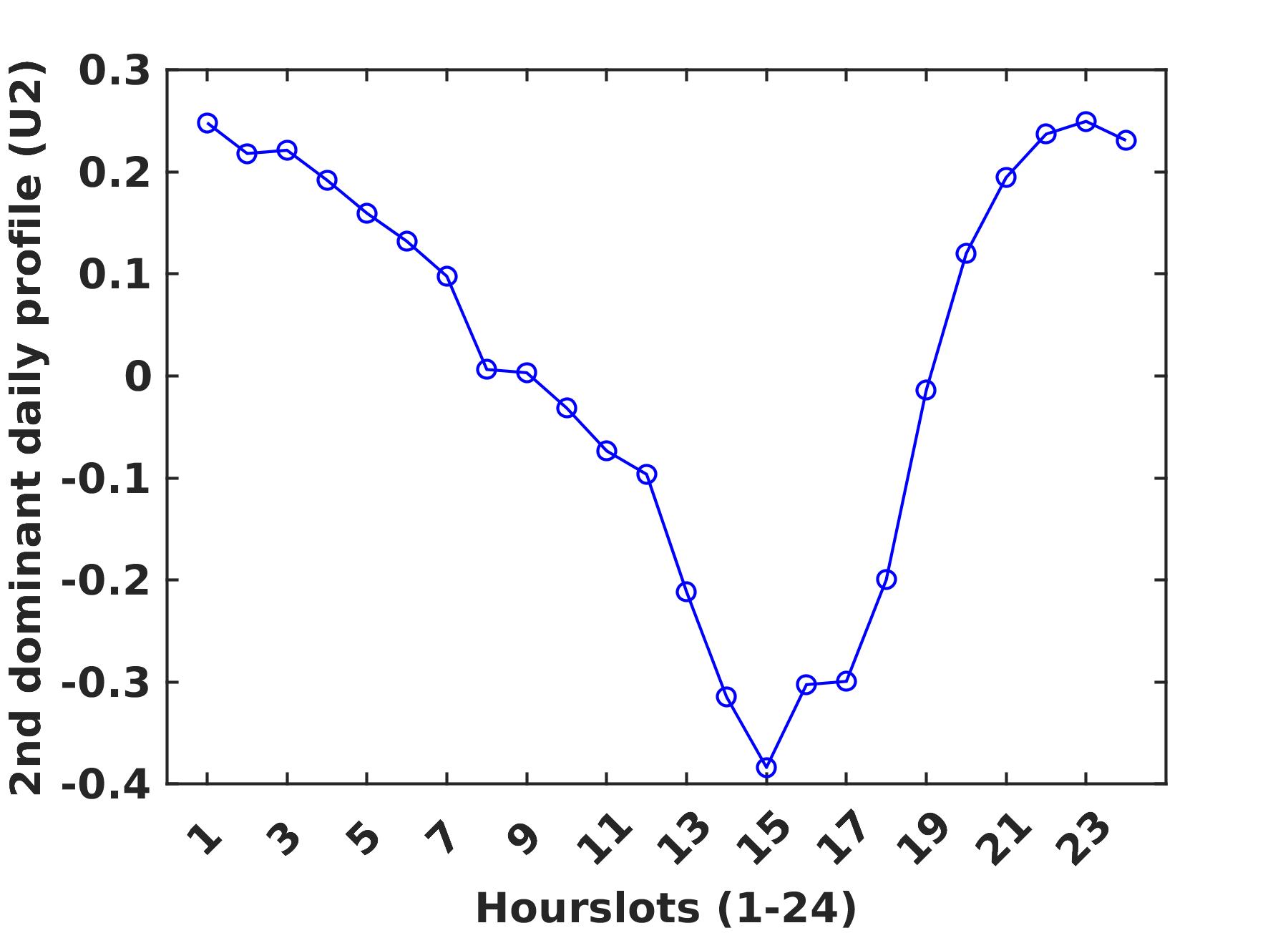}
    \includegraphics[width=7.5cm,height=3.3cm]{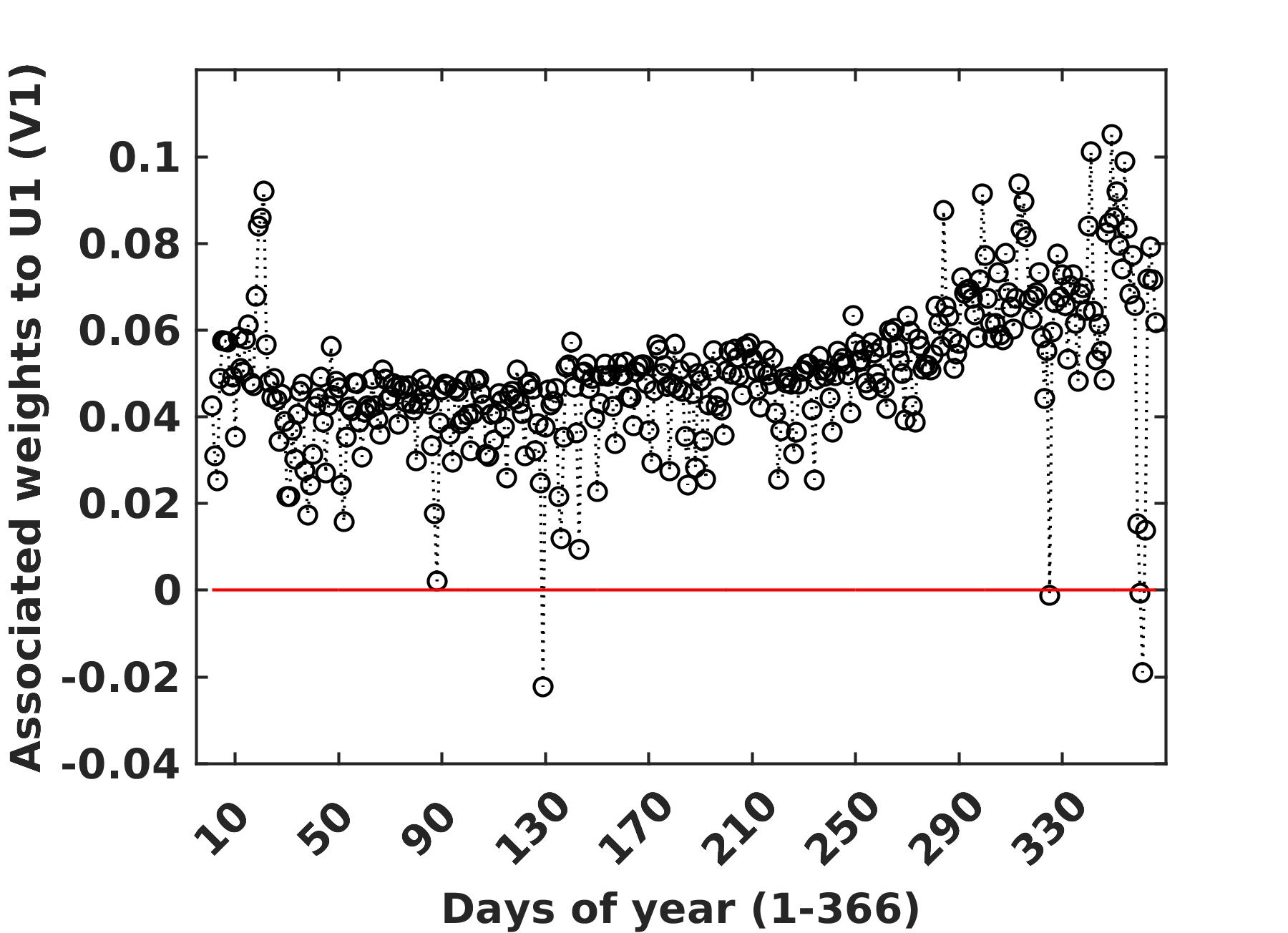}
     \includegraphics[width=7.5cm,height=3.3cm]{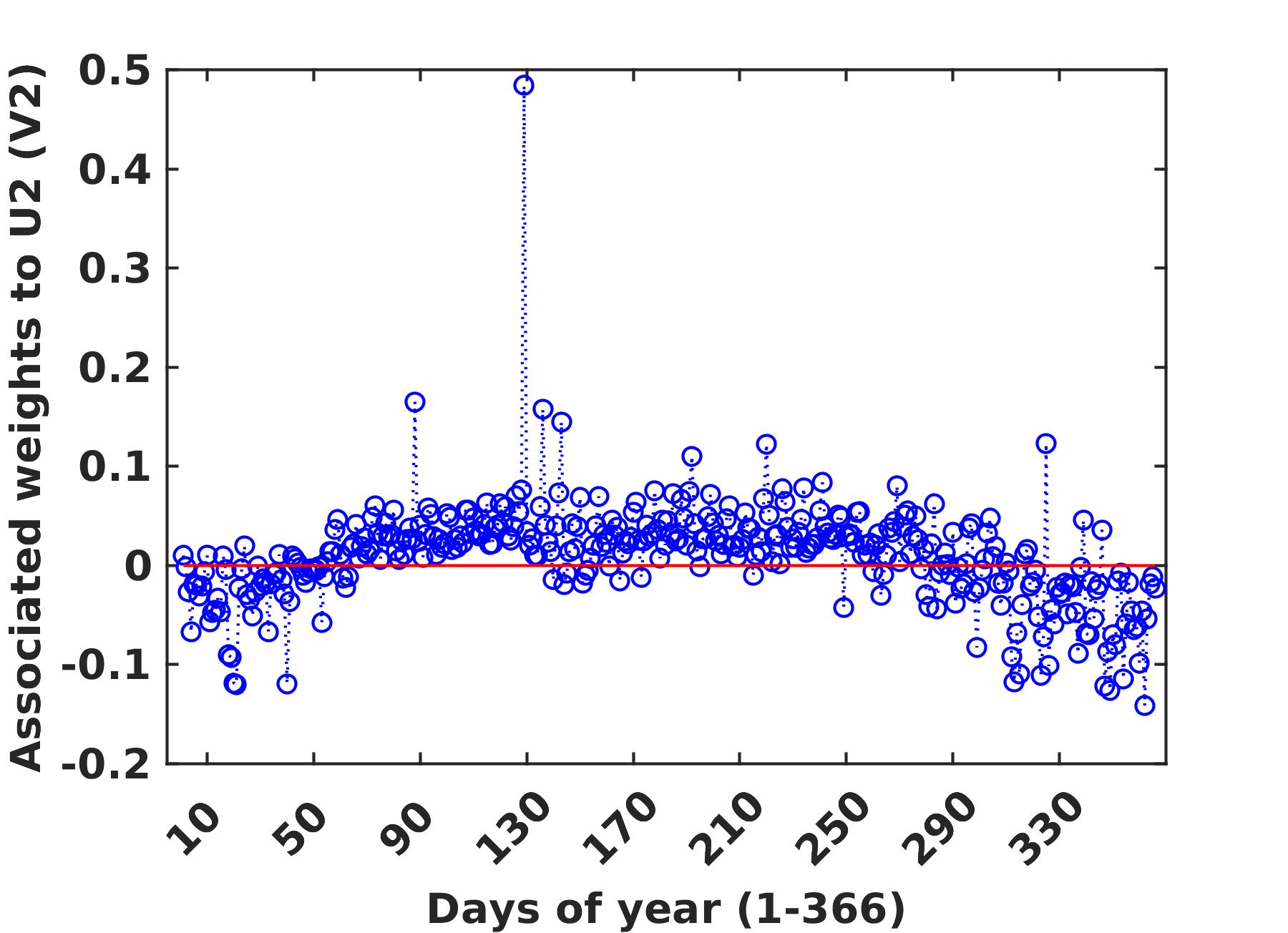}
    \caption{From top to bottom: The first two panels are the first ($U_1$) and second ($U_2$) dominant profiles of the price profiles in 2016; the third and the last one are their corresponding amplitudes ($V_1$ and $V_2$) throughout the year.}
    \label{fig:U1V1U2V2}
\end{figure}

\begin{figure}[]
    \centering
    \includegraphics[width=7.5cm,height=3.3cm]{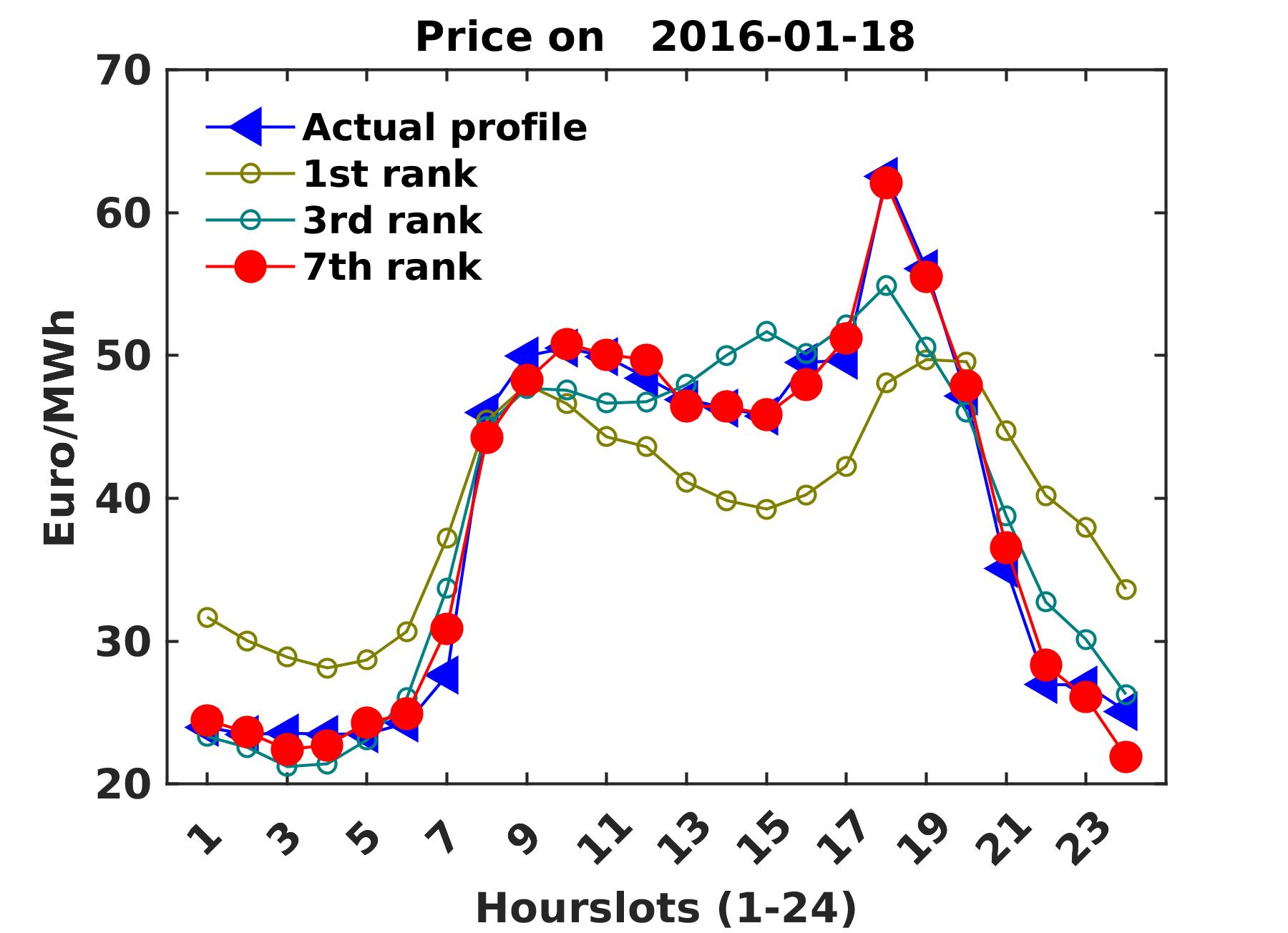}
    \caption{Low rank approximation of actual data (one particular day, Jan. 18, blue).
    Including up to 7 SVD components yields the rank-7 approximation (bold red).  Lower rank approximations are also shown. }
    \label{fig:main_20170302_svd_xrank_reconstruction}
\end{figure}

Another advantage of representing the time series as matrices is the possibility of using the matrix decomposition techniques to analyze the structure of the data. 
In this paper we will focus on the well-known singular value decomposition (SVD) method which states that an arbitrary $h \times d$  matrix $A$ of rank $r \leq min(h,d)$ can be factored 
as: 
\begin{equation}
    A  =USV^T = \sum\limits_{k = 1}^r \sigma_kU_k V_k^T 
    \label{eq:svd}
\end{equation}
where $U \in {\cal O }(h)$  and $V  \in {\cal O }(d)$  are orthonormal matrices,
(with $U_k$ and $V_k$ denoting the $k^{th}$ column of $U$ and $V$, respectively) and $S$ is an $h\times d$ matrix for which the only strictly positive elements $\sigma_k$ (so-called singular values) are placed on the main diagonal (see~{\cite{baker2005singular}}).
The importance of this decomposition lies in the fact that truncating the expansion in the right hand side of~\eqref{eq:svd} after the $p^{th}$ term yields the best  approximation of the original matrix $A$ by a matrix of (lower) rank $p$ (see~{\cite{golub1970singular}}). 
%
%
\begin{equation}
     A_p = \arg\min\limits_{rank(R) = p} \n A-R \n 
\end{equation}
where the norm $ \n \cdot \n $ can be either the Frobenius or spectral $L_2$ norm. 
Translating these results back to the original time series, we see that the columns of $U\in {\cal O}(24)$ represent daily profiles, whereas the columns of $V\in {\cal O}(366)$ furnish the corresponding amplitudes 
(one for each day). Fig.~\ref{fig:U1V1U2V2} shows a concrete illustration for the price data: 
the first column $U_1$ is 
depicted in the top panel and provides an overall daily profile (obtained as a weighted average). 
The expected price peaks in the morning and early evening are clearly discernible. The 
corresponding amplitudes (one for each day) are given by the column $V_1$ and shown in the 3rd panel. 
Some days with exceptionally low or high prices are clearly visible.\\
A rank-1 approximation of the original time series would therefore be obtained by taking the averaged daily profile $U_1$ in the top panel and scaling it up or down using the 366 values in $V_1$ (depicted in the third panel). 
Notice that in this first approximation, each day has the same profile, only the amplitudes  
change from one day to another.  
The values of $U_2$ (depicted in panel 2) specify a first correction to $U_1$, with the 
corresponding amplitudes for this correction specified in $V_2$ (displayed in panel 4). 
This correction means that for any day for which the corresponding $V_2$ coefficient is positive 
(almost every summer day) will have a lower price value between 11h and 18h than would have expected based on the (weighted) annual average $U_1$.  
Adding additional terms in the SVD expansion improves the approximation. 
This is illustrated in Fig.~\ref{fig:main_20170302_svd_xrank_reconstruction} 
where the actual data for one typical day (18 Jan. 2016) are shown in conjunction with low-rank approximation 
up to rank 7.
\subsection{Structure-preserving smoothing}
\begin{figure}[]
    \centering
    \includegraphics[width=7.5cm,height=3.3cm]{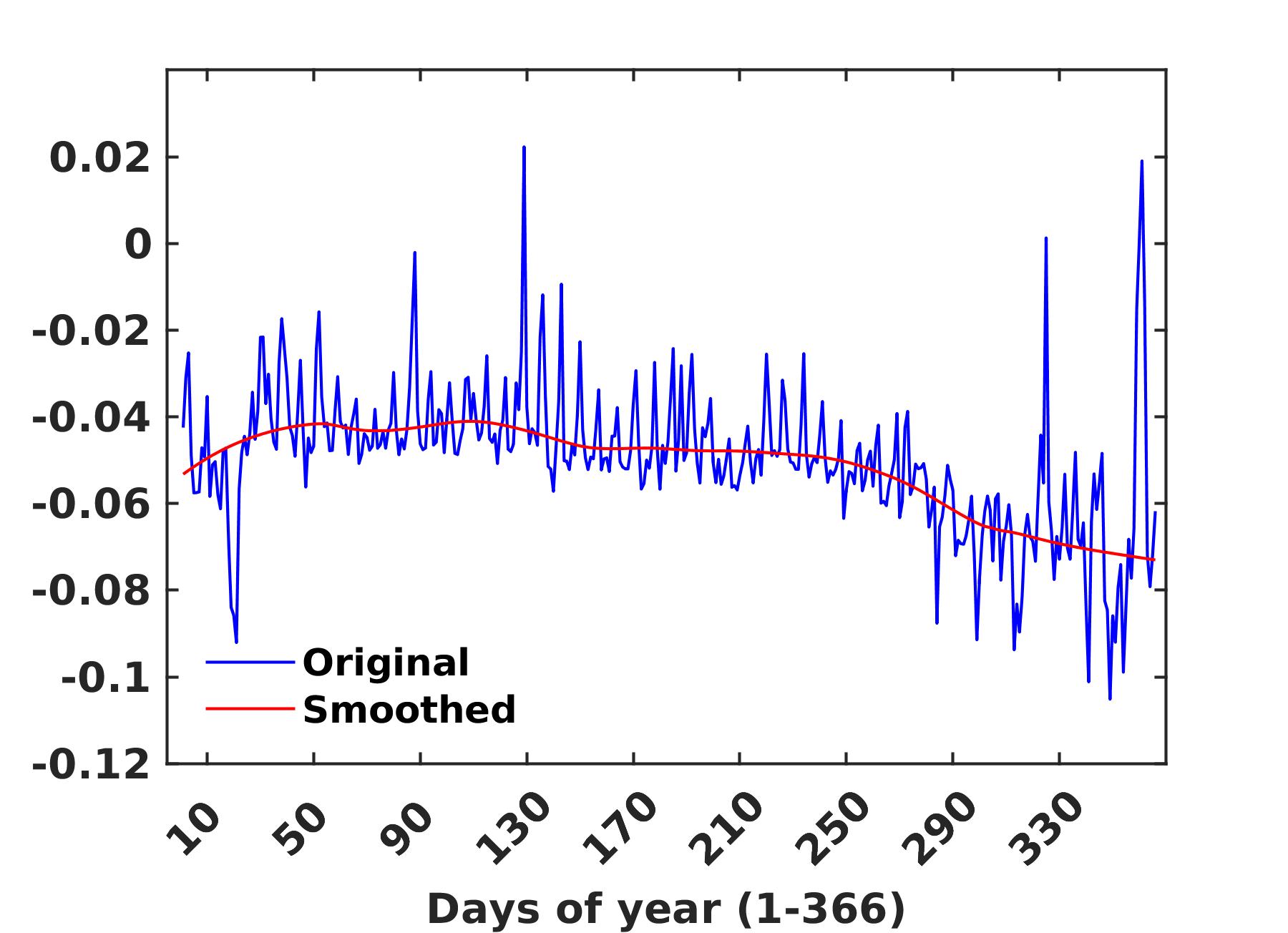}
    \caption{An example of a smoothed $V_1$ (corresponding to the most dominant singular value) used in reconstruction of the 2nd derivative of the price data.}
    \label{fig:main_20170302_svd_smoothed_V}
\end{figure}

\begin{figure*}[!h]
     \begin{center}{%
}%
        \subfigure[Intra-day 2nd derivative of price profiles]{%
            \label{fig:main_20170317_gyy_pr}
            \includegraphics[width=5cm,height=3.3cm]{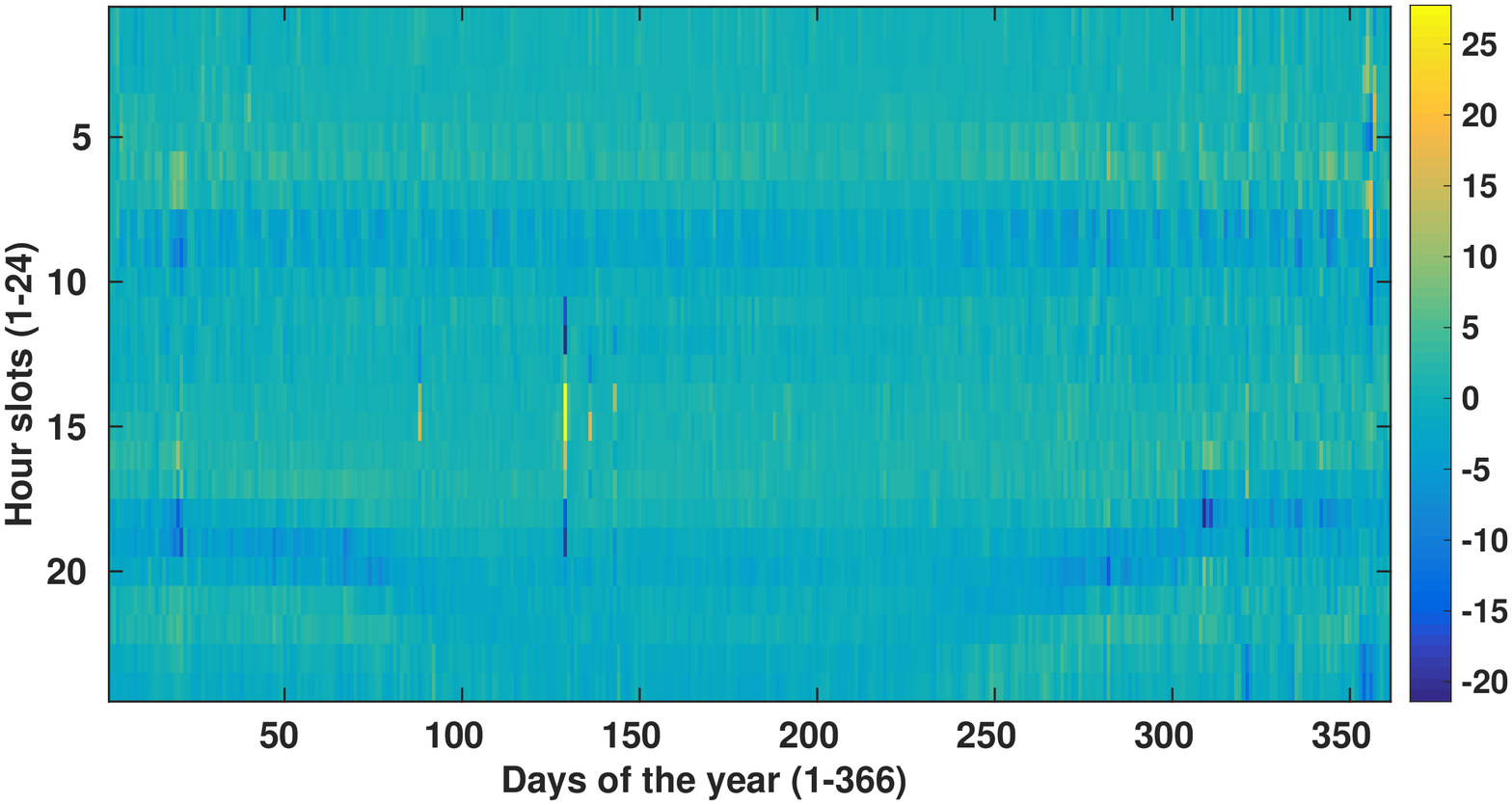}
        } 
        \subfigure[1st rank reconstruction of Fig.~\ref{fig:main_20170317_gyy_pr}]{%
           \label{fig:main_20170317_gyy_pr_rank1}
           \includegraphics[width=5cm,height=3.3cm]{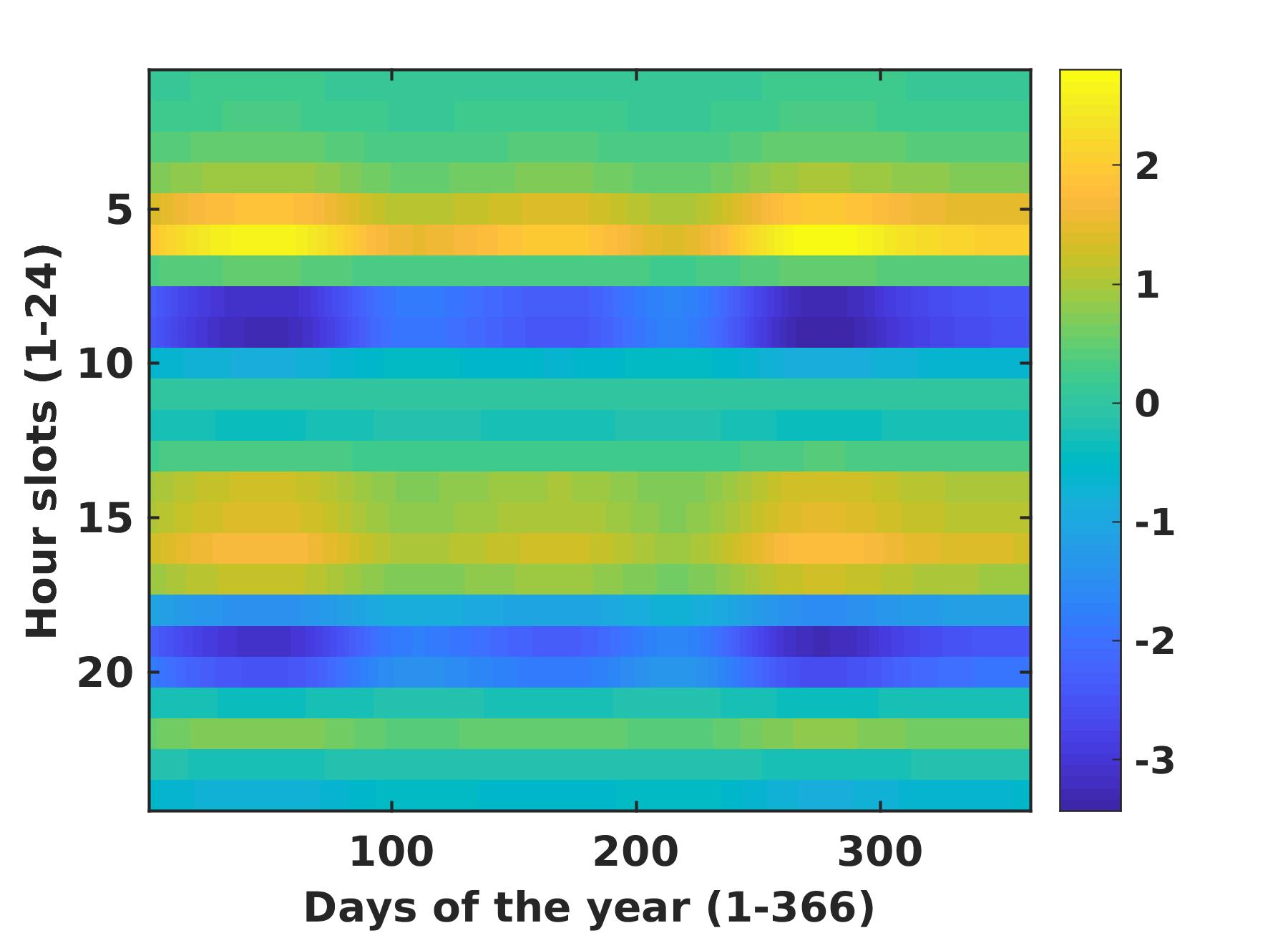}
        } 
         \subfigure[7th rank reconstruction of Fig.~\ref{fig:main_20170317_gyy_pr}]{%
           \label{fig:main_20170317_gyy_pr_rank7}
           \includegraphics[width=5cm,height=3.3cm]{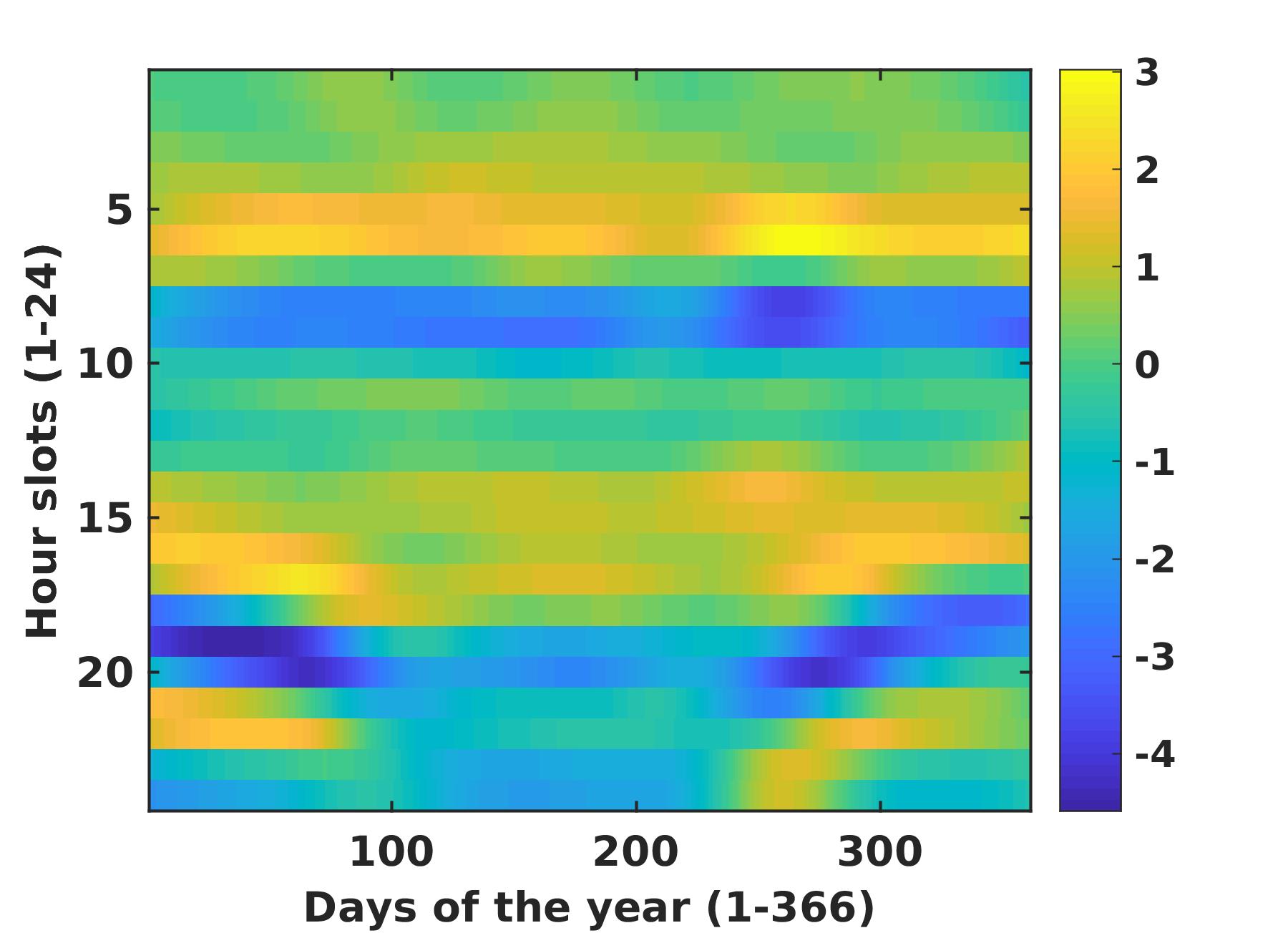}
        } \\
               \subfigure[Intra-day 2nd derivative of solar feed-in]{%
            \label{fig:main_20170317_gyy_sol}
            \includegraphics[width=5cm,height=3.3cm]{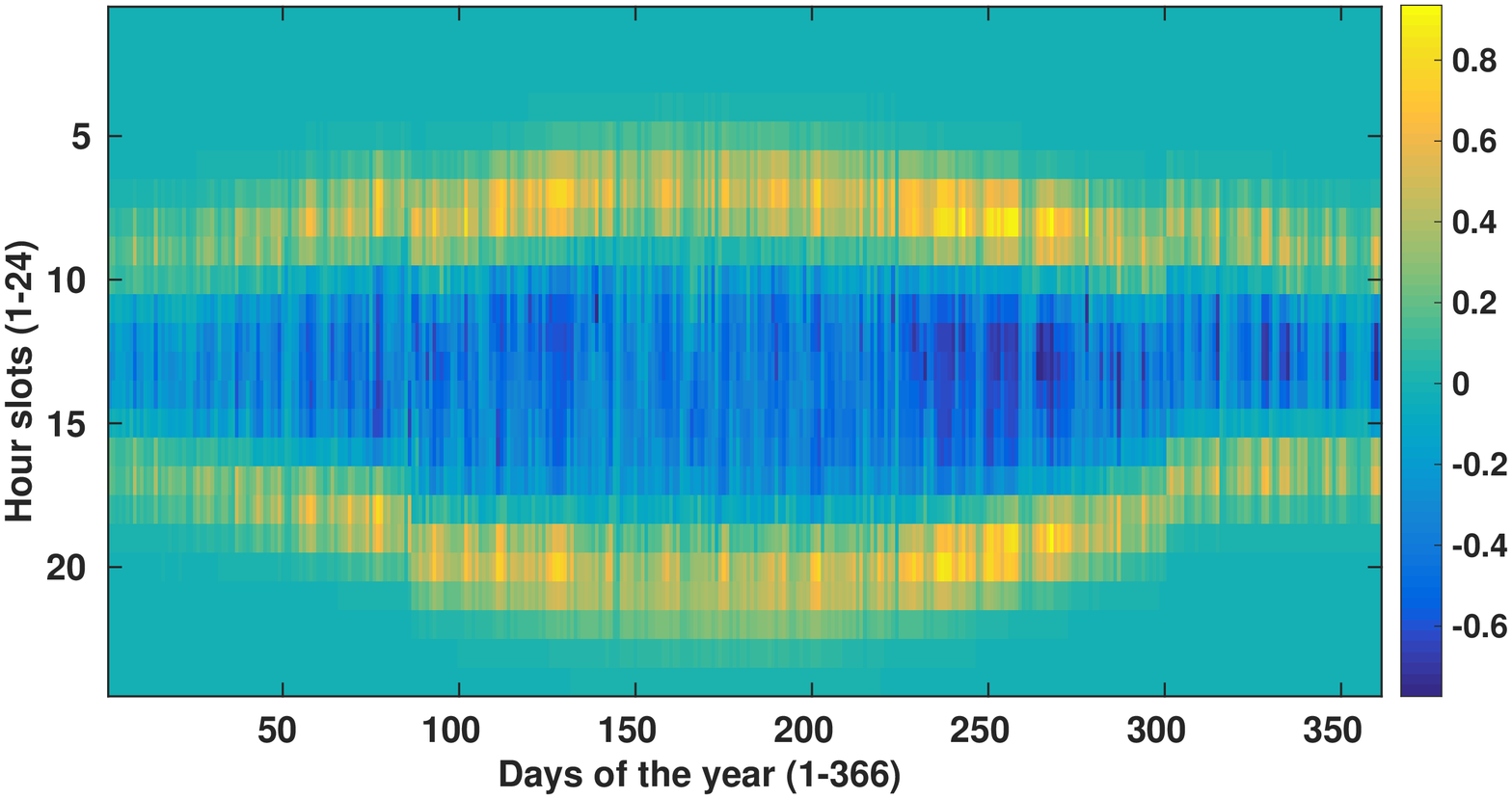}
        } 
        \subfigure[1st rank reconstruction of Fig.~\ref{fig:main_20170317_gyy_sol}]{%
           \label{fig:main_20170317_gyy_sol_rank1}
           \includegraphics[width=5cm,height=3.3cm]{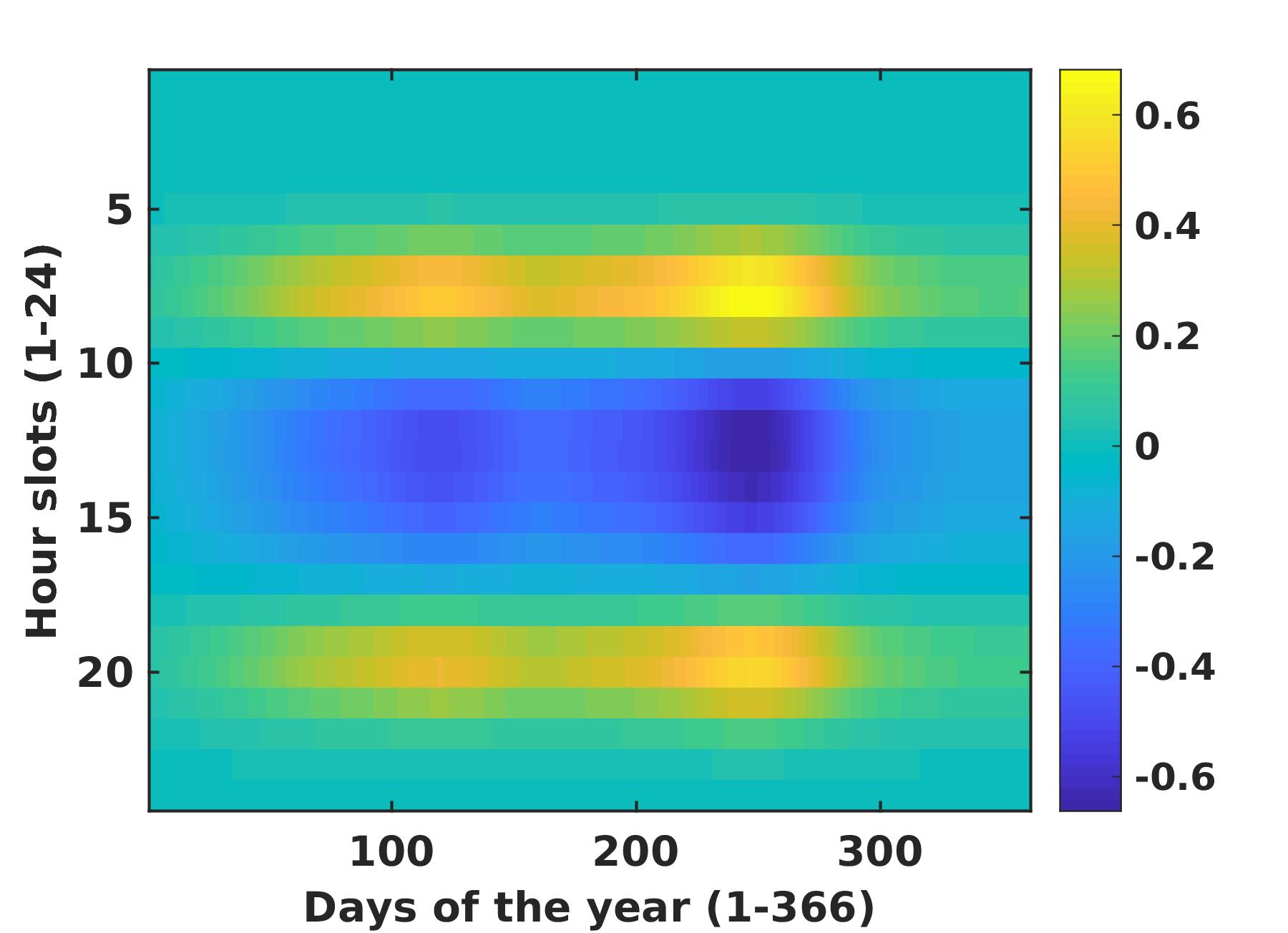}
        } 
         \subfigure[7th rank reconstruction of Fig.~\ref{fig:main_20170317_gyy_sol}]{%
           \label{fig:main_20170317_gyy_sol_rank7}
           \includegraphics[width=5cm,height=3.3cm]{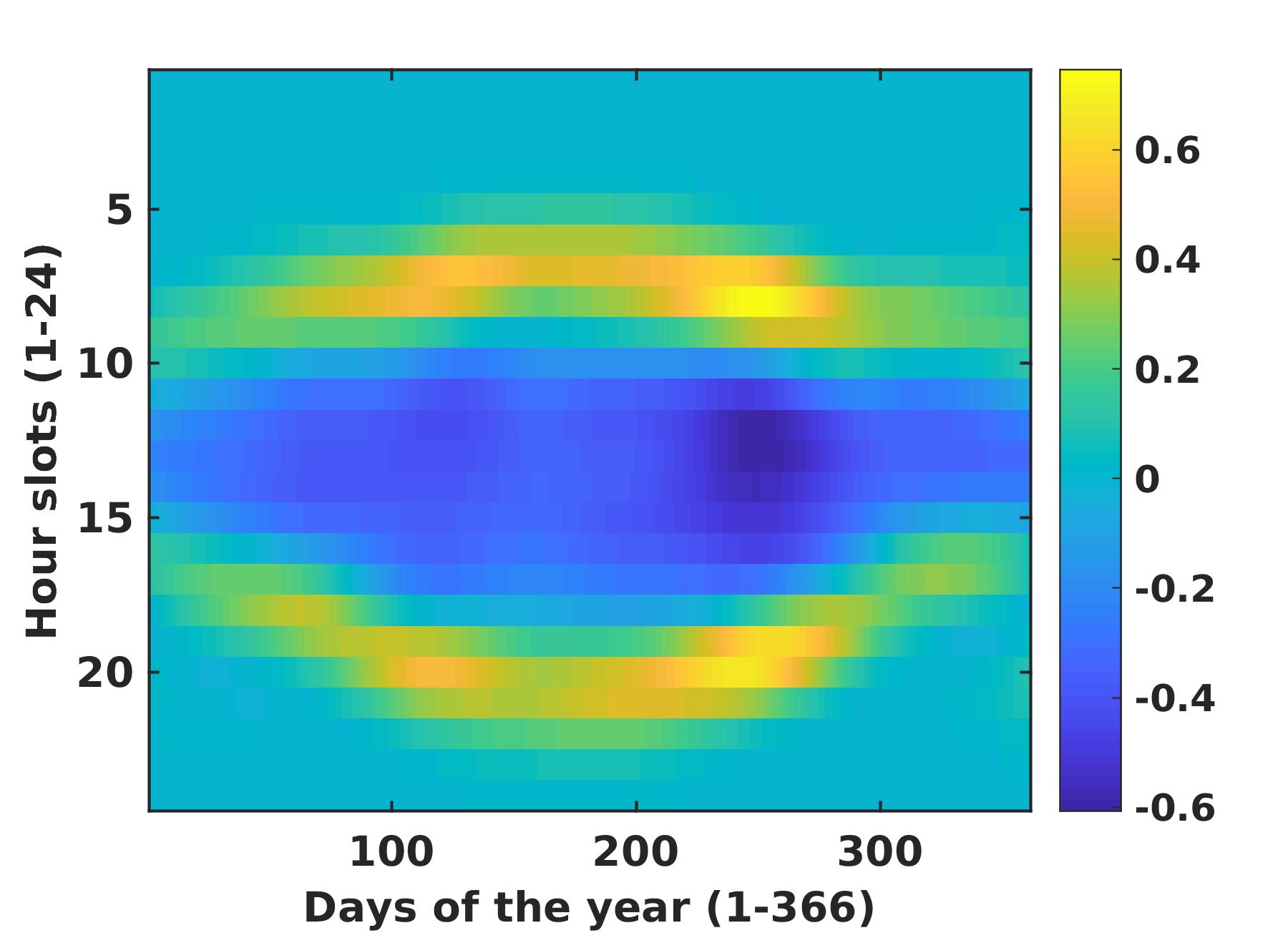}
        } \\
                \subfigure[Intra-day 2nd derivative of wind feed-in]{%
            \label{fig:main_20170317_gyy_wd}
            \includegraphics[width=5cm,height=3.3cm]{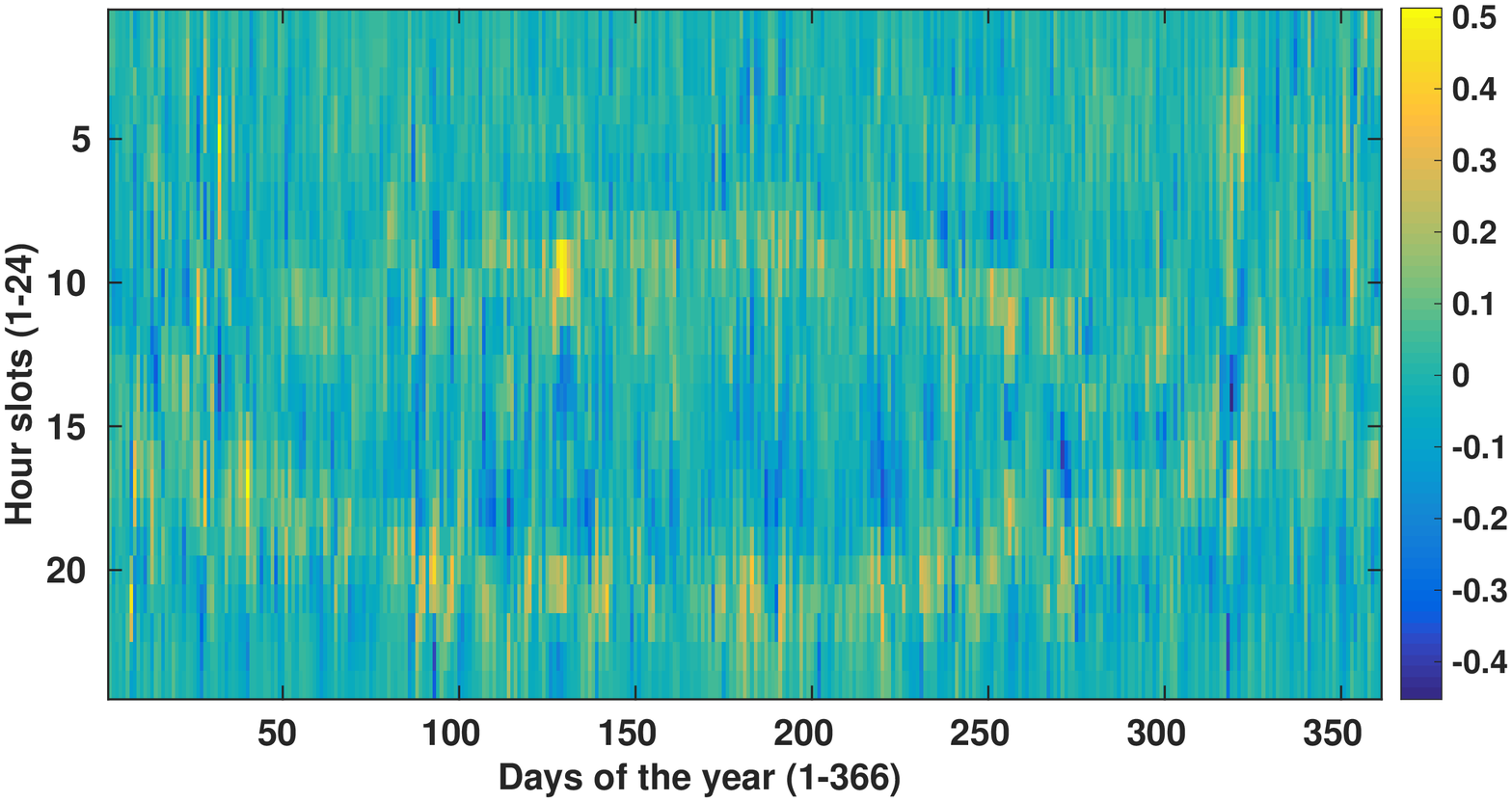}
        } 
        \subfigure[1st rank reconstruction of Fig.~\ref{fig:main_20170317_gyy_wd}]{%
           \label{fig:main_20170317_gyy_wd_rank1}
           \includegraphics[width=5cm,height=3.3cm]{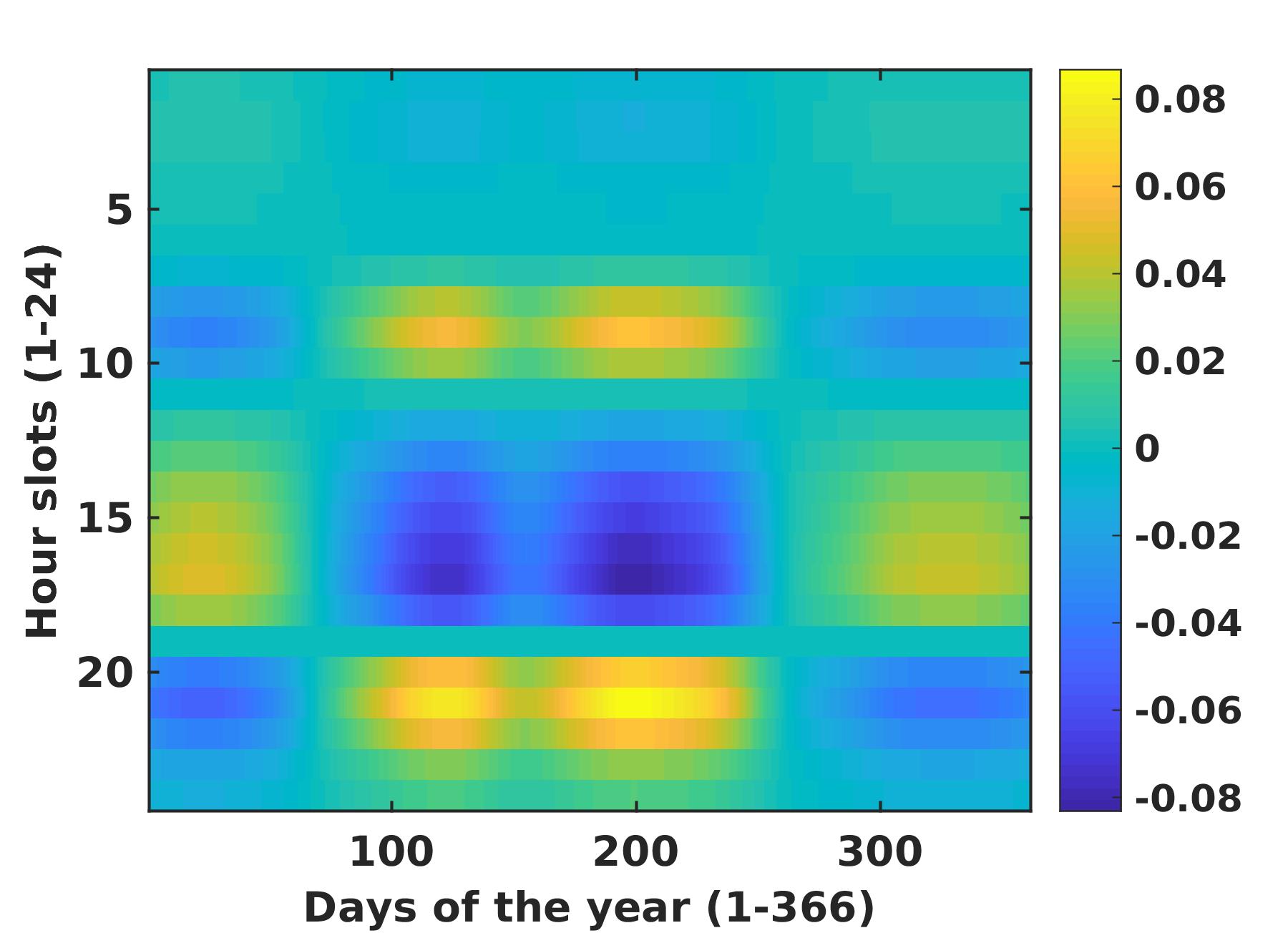}
        } 
         \subfigure[7th rank reconstruction of Fig.~\ref{fig:main_20170317_gyy_wd}]{%
           \label{fig:main_20170317_gyy_wd_rank7}
           \includegraphics[width=5cm,height=3.3cm]{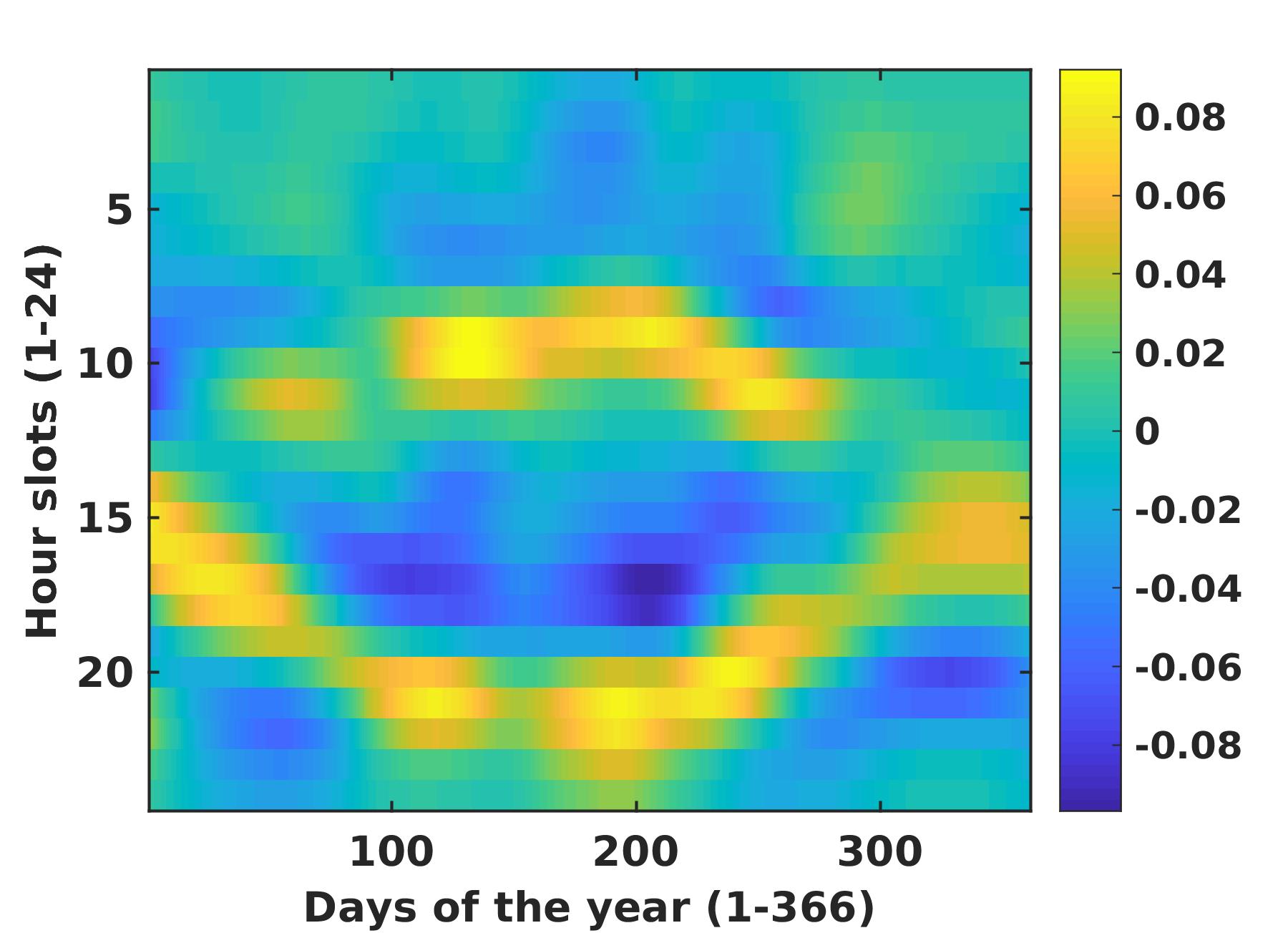}
        }\\
                \subfigure[Intra-day 2nd derivative of load profiles]{%
            \label{fig:main_20170317_gyy_ld}
            \includegraphics[width=5cm,height=3.3cm]{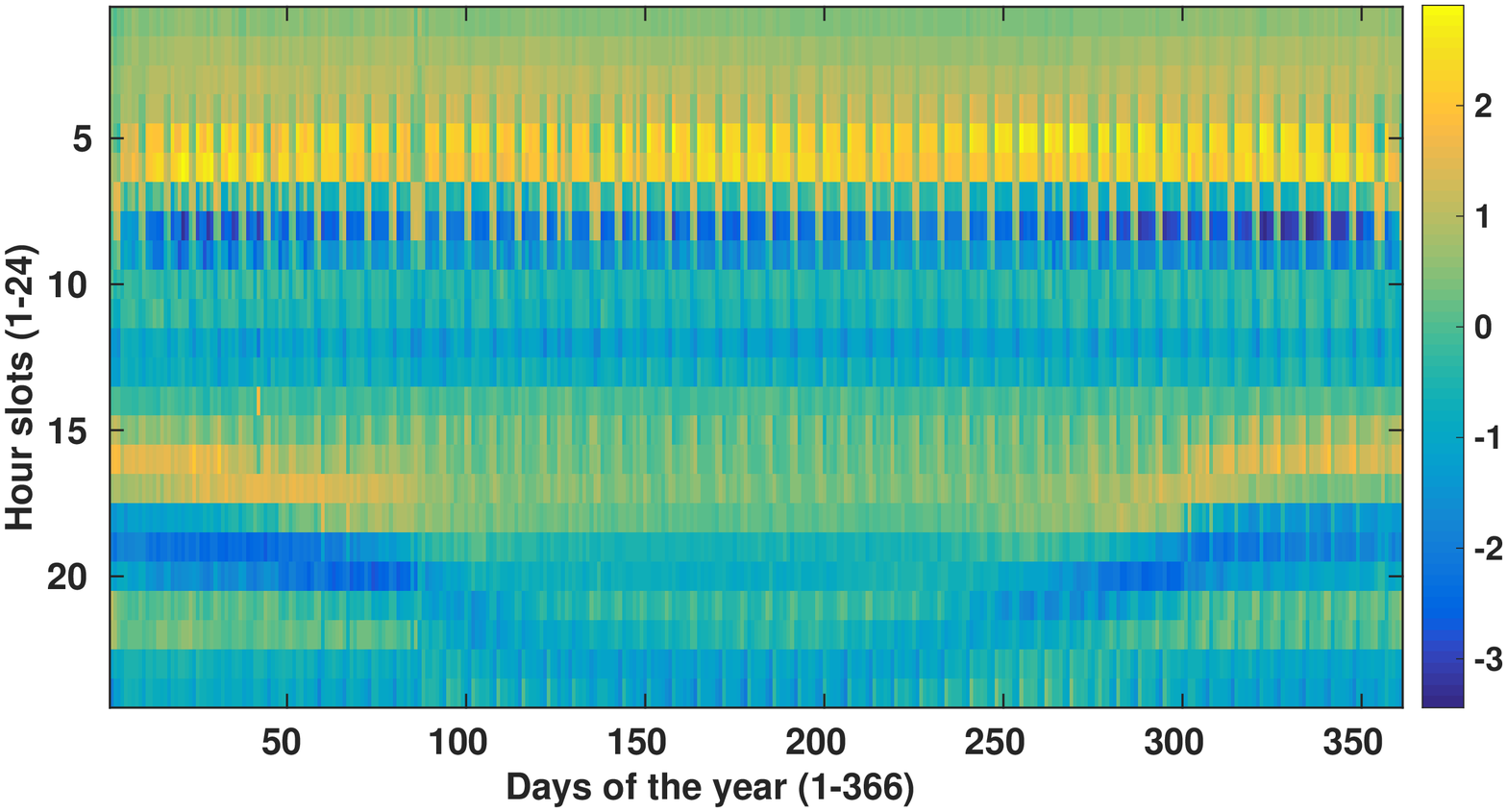}
        } 
        \subfigure[1st rank reconstruction of Fig.~\ref{fig:main_20170317_gyy_ld}]{%
           \label{fig:main_20170317_gyy_ld_rank1}
           \includegraphics[width=5cm,height=3.3cm]{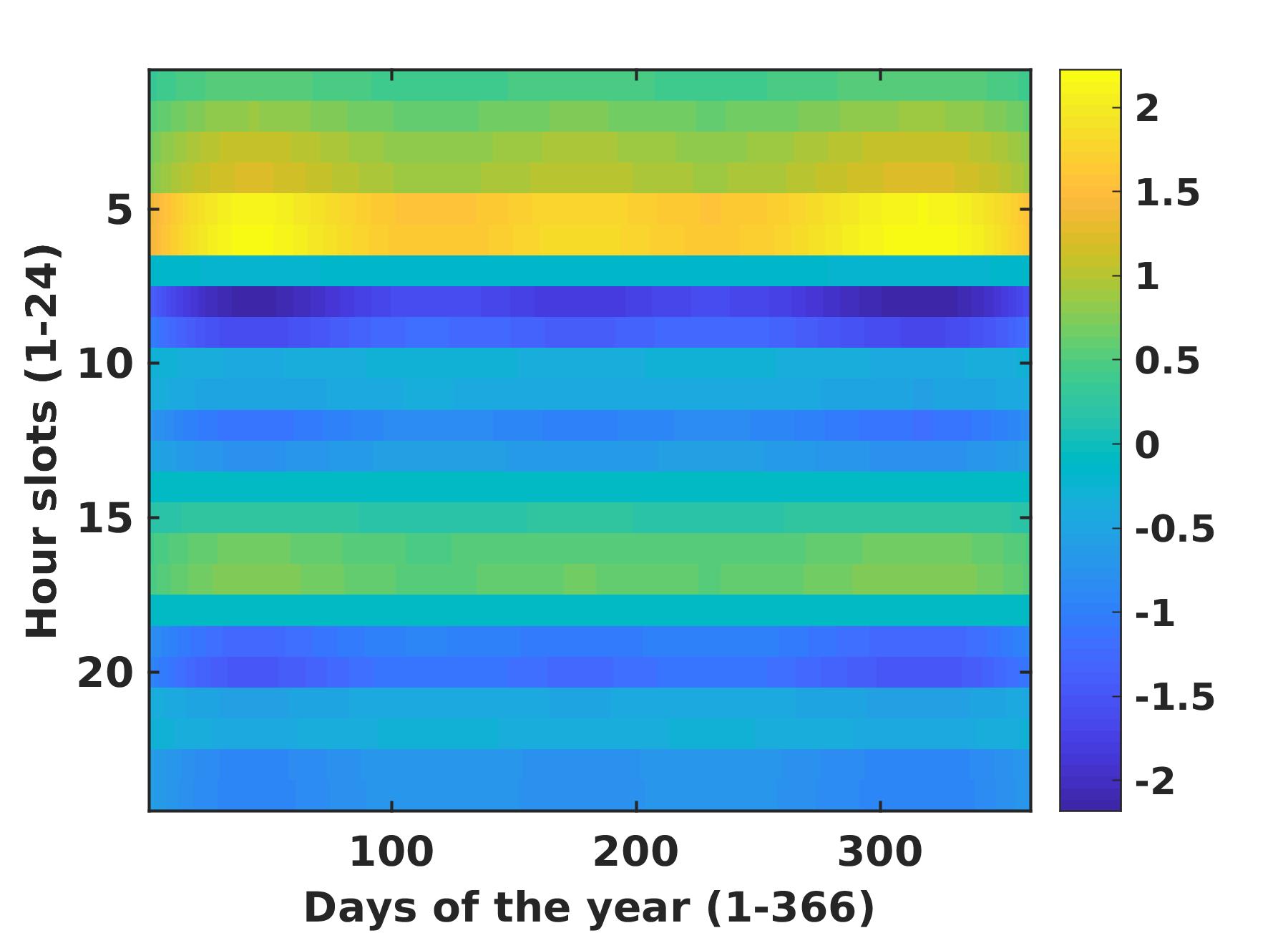}
        } 
         \subfigure[7th rank reconstruction of Fig.~\ref{fig:main_20170317_gyy_ld}]{%
           \label{fig:main_20170317_gyy_ld_rank7}
           \includegraphics[width=5cm,height=3.3cm]{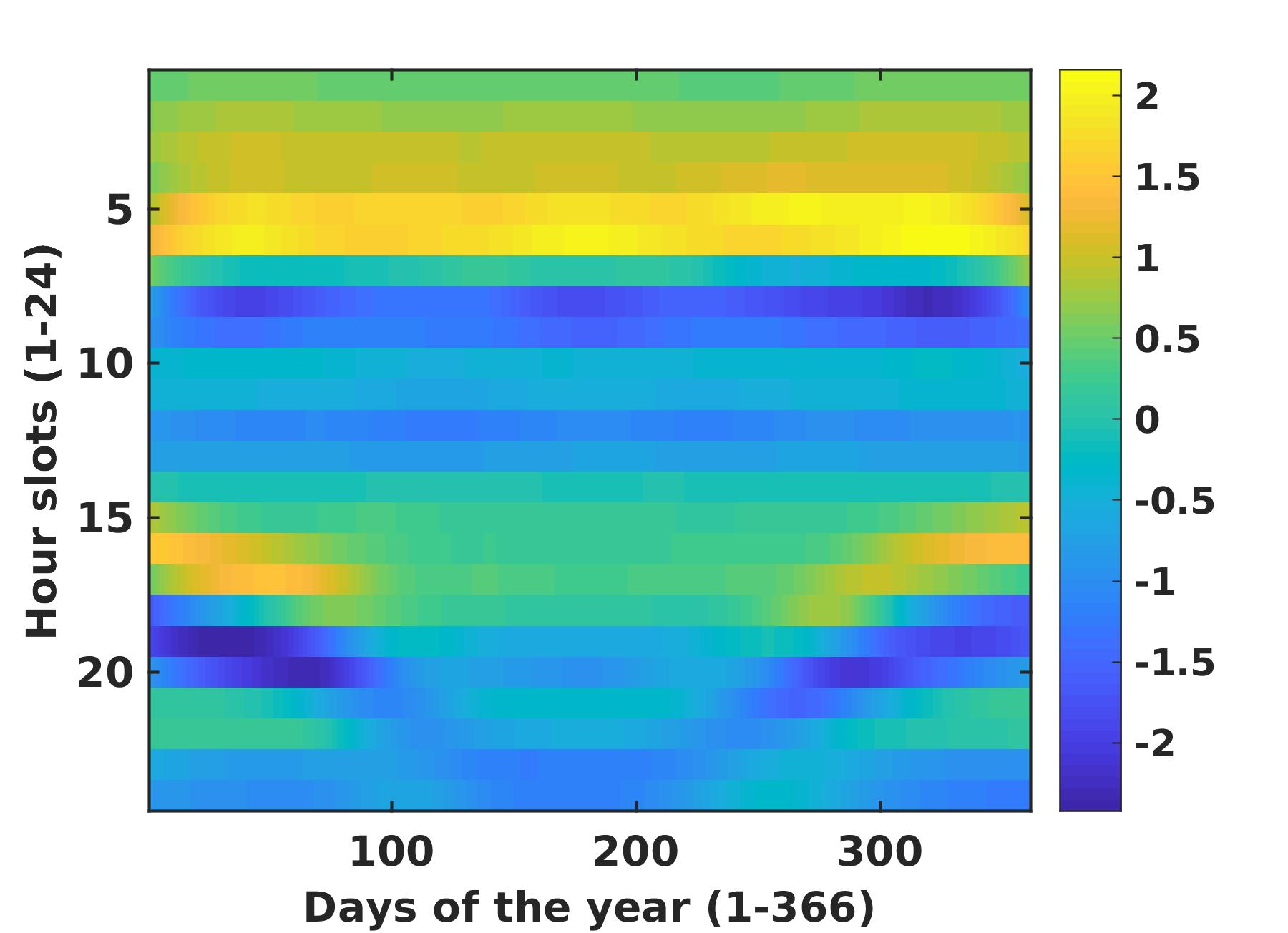}
        }\\
                \subfigure[Intra-day 2nd derivative of the traded quantity ]{%
            \label{fig:main_20170317_gyy_qn}
            \includegraphics[width=5cm,height=3.3cm]{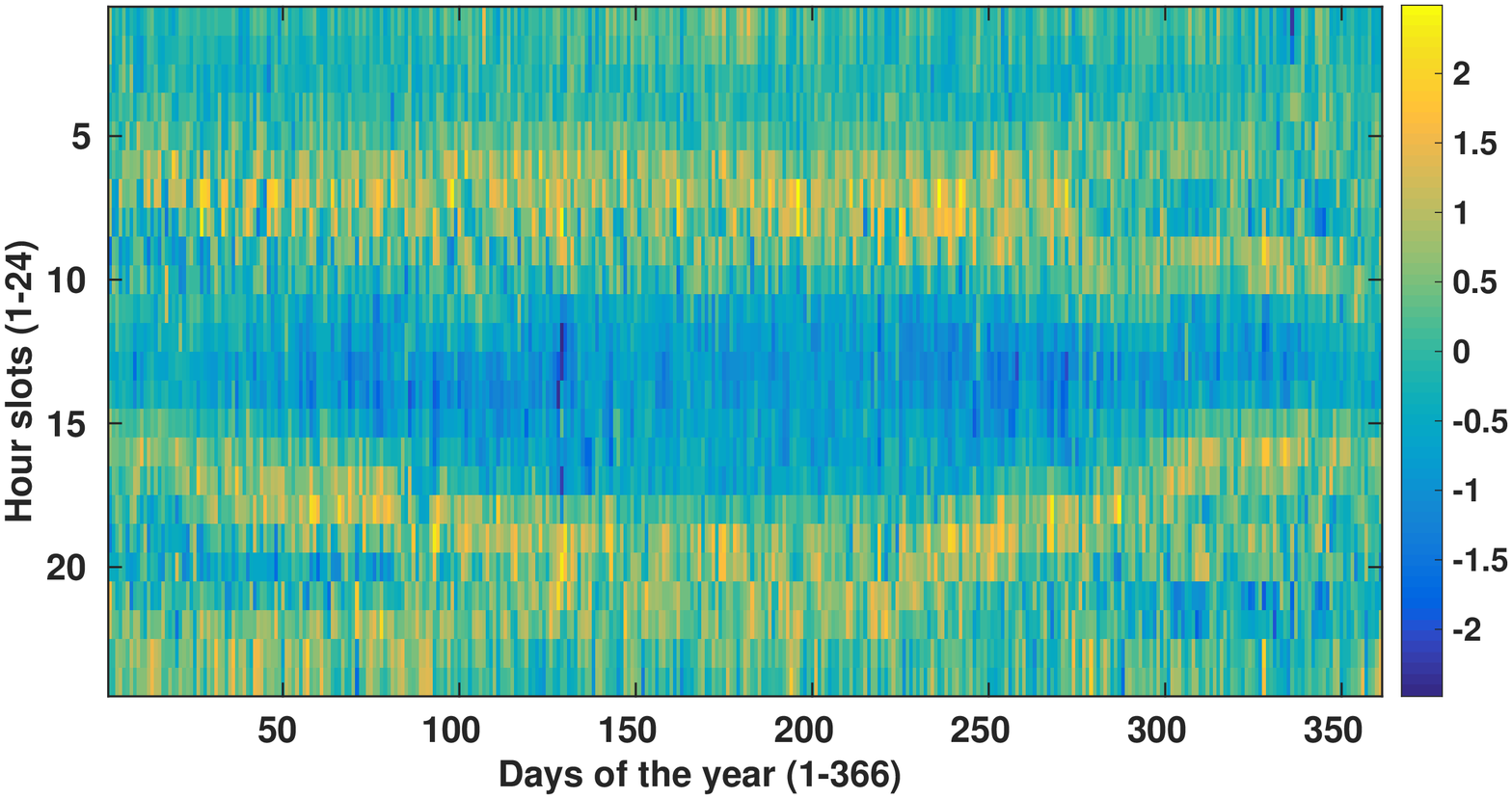}
        } 
        \subfigure[1st rank reconstruction of Fig.~\ref{fig:main_20170317_gyy_qn}]{%
           \label{main_20170317_gyy_qn_rank1}
           \includegraphics[width=5cm,height=3.3cm]{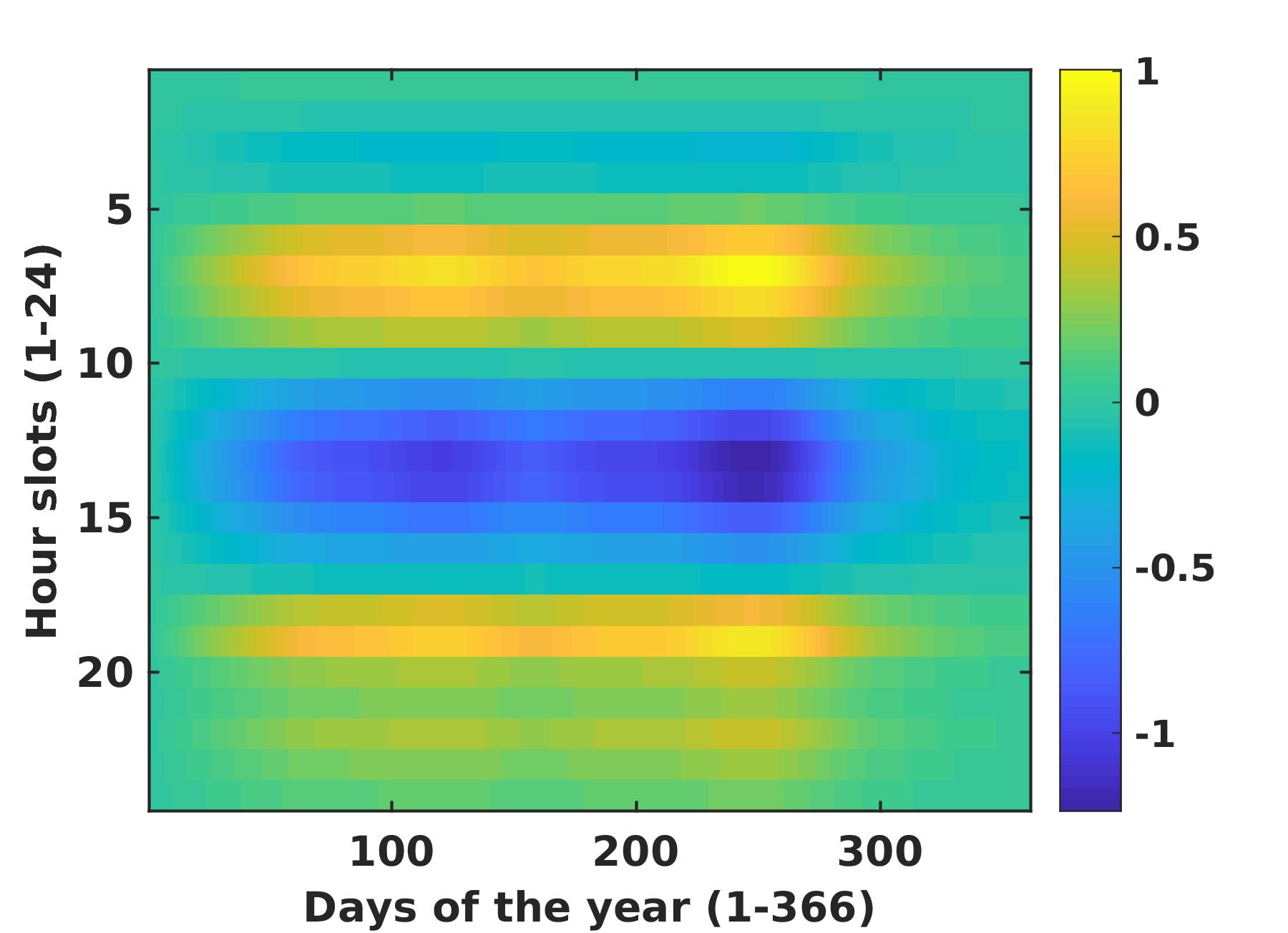}
        } 
         \subfigure[7th rank reconstruction of Fig.~\ref{fig:main_20170317_gyy_qn}]{%
           \label{fig:main_20170317_gyy_qn_rank7}
           \includegraphics[width=5cm,height=3.3cm]{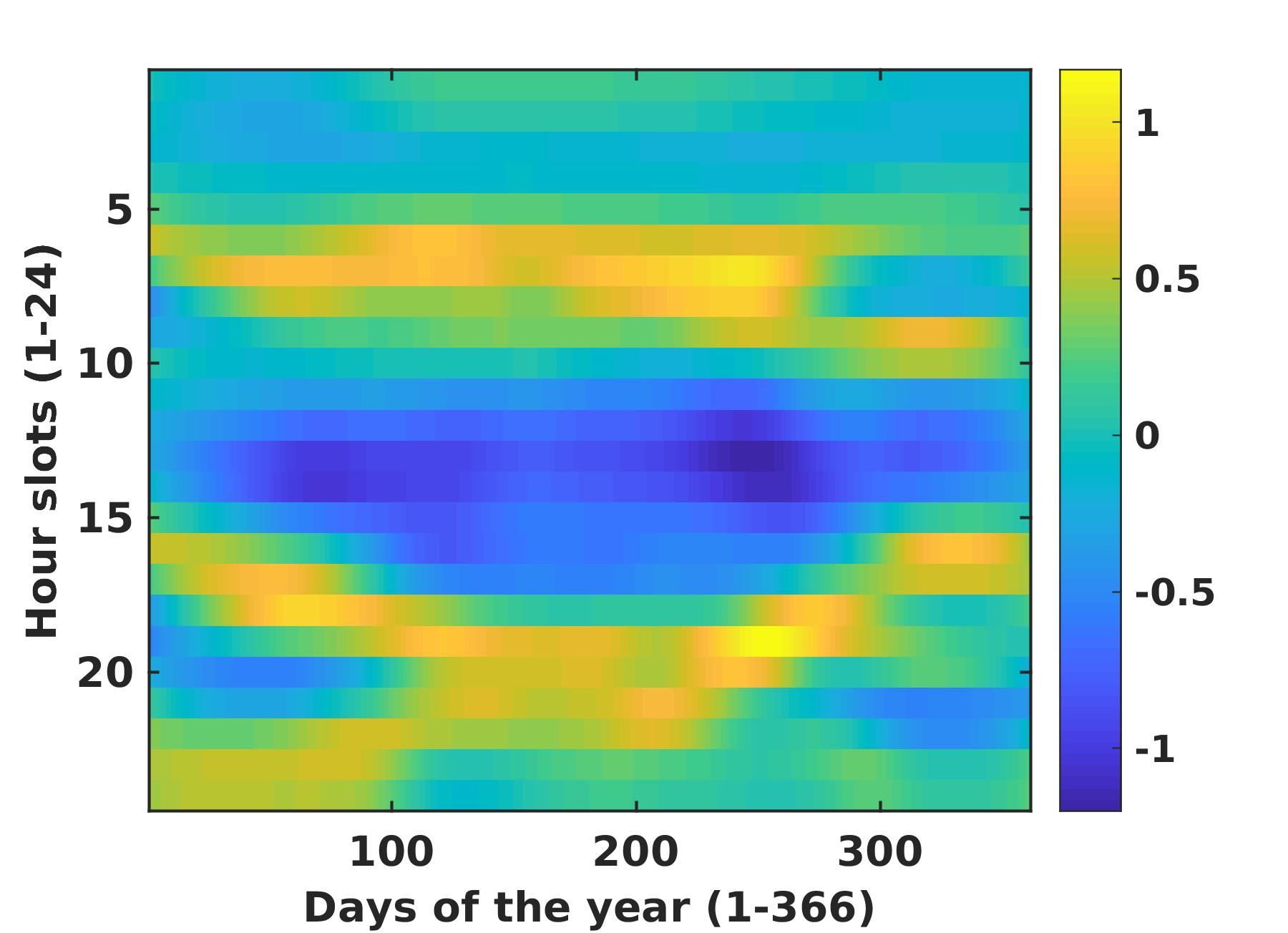}
        }
    \end{center}
    \vspace*{-2mm}
    \caption{%
   Left: The intra-day 2nd derivative (concavity) of the  German day-ahead market data in 2016, 
   for (top to bottom) price, solar  and wind feed-in, load and traded quantity. 
   The underlying trends have been magnified using rank-1 (middle) 
   and rank-7 (right) reconstruction. Clearly, the rank-7 reconstruction yields an acceptable 
   approximation of the original images on the left.}
   \label{fig:main_20170302_svd}
\end{figure*}

\begin{figure}[]
    \centering
    \includegraphics[width=7.5cm,height=3.7cm]{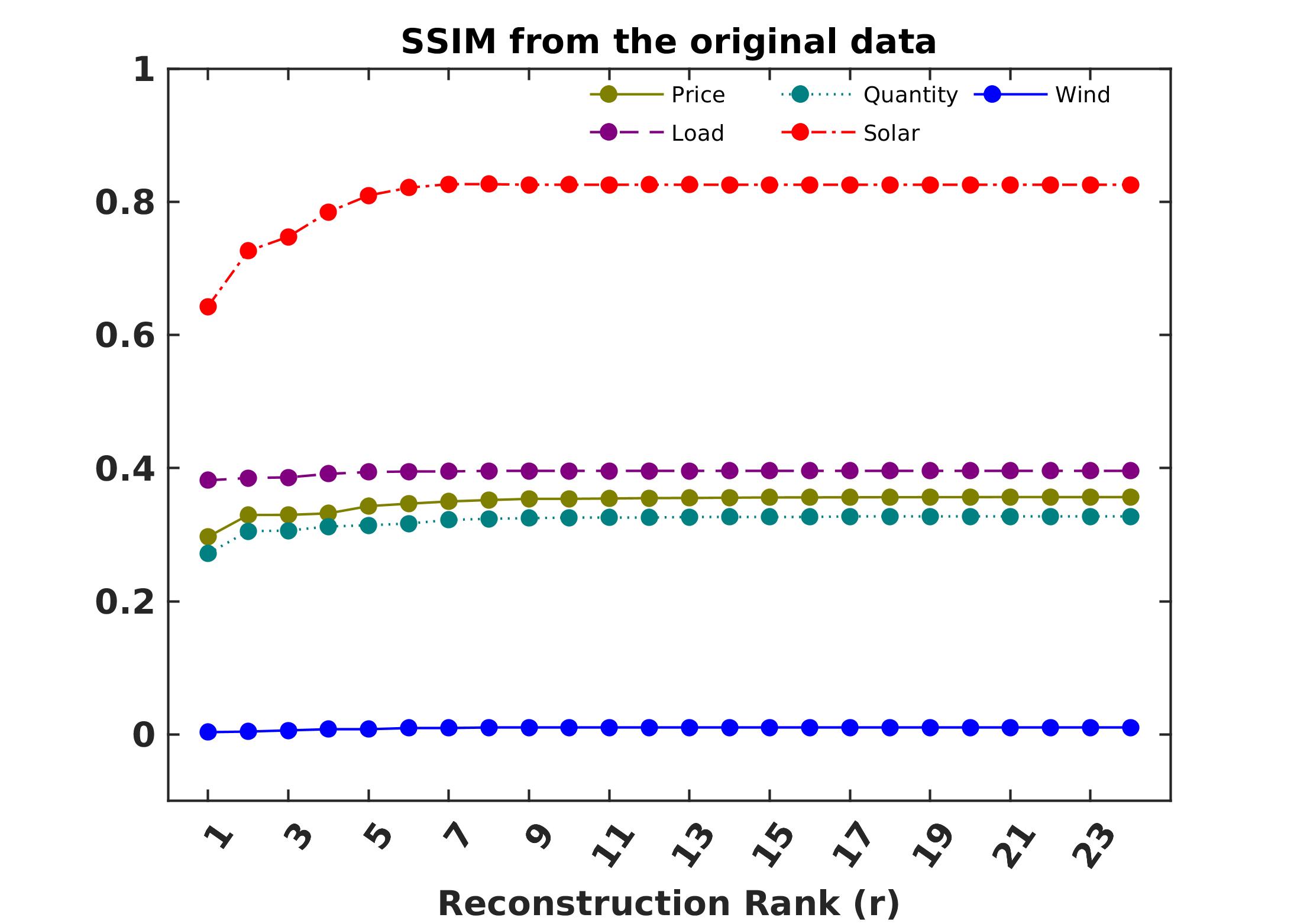}
    \caption{The evolution of {SSIM} of the reconstructed matrix, $A_p$ (for $p=1,...,24$) from the original 2nd derivative matrix $A$. Thus rank-7 was considered an appropriate choice in our experiments.}
    \label{fig:main_20170306_svd_why_rank_fig123}
\end{figure}

As mentioned earlier, recasting the time series as images allows us to visually integrate subtle patterns in the data. 
Now, we indicate how the SVD factorization suggests a straightforward method to smooth the time-series in such a way that the overall structure is preserved.
Recall that the $V$ columns determine the amplitudes of every daily-profile.
By smoothing these profiles we eliminate most of the inter-day variation without affecting the overall structure. 
For the data at hand, the smoothing was based on robust local regression (RLOESS) 
but alternative approaches would be equally valid. 
The local regression smoothing method was used to alleviate the effect of outliers, 
while preserving the general trend in data.
Fig.~\ref{fig:main_20170302_svd_smoothed_V} shows an example of the smoothed $V_1$ applied 
in the calculation of $A_p$.
Using the smoothed version of the $V$ vectors in the low rank reconstruction of data, yields the 
images depicted in Fig.~\ref{fig:main_20170302_svd}. 
The panels in the left hand column show the original 2nd derivative profiles for the various quantities of interest (price, solar and wind feed-in, load and traded quantity). 
The two adjacent columns show the same data, but this time using two different smoothed, low-rank approximations: a rough rank-1 approximation (middle) and a much more accurate rank-7 approximation, which 
will be used in the regression analysis (Section~\ref{sec:results}).  
In addition to appearing visually unbiased
the choice for rank-7 is also based on the structural similarity index (SSIM), 
typically used for measuring image quality.     
{SSIM} assesses the deviation of $A_p$ from the original matrix $A$ by comparing their corresponding local means, standard deviations, and cross-covariance matrices.
For detailed description, readers are referred to~{\cite{matlab_ssim}}.
As can be seen in Fig.~\ref{fig:main_20170306_svd_why_rank_fig123}, in terms of {SSIM}, the approximations keep improving up till $p=7$  after which it levels off. As a consequence, we select rank-7 as the appropriate approximation. 
It is worth noting that this figure is in line with our expectation regarding volatility of the various quantities: 
wind results in the lowest similarity, while solar feed-in agrees best with the approximation.
%
\section{Results and Conclusions} \label{sec:results}
\begin{figure}[]
    \centering
    \includegraphics[width=7.5cm,height=2.5cm]{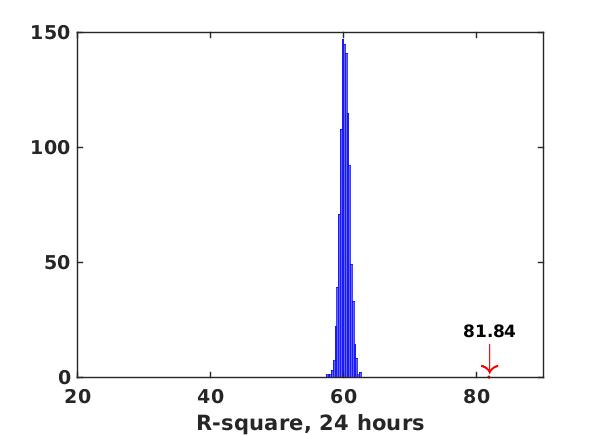}
    \includegraphics[width=7.5cm,height=2.5cm]{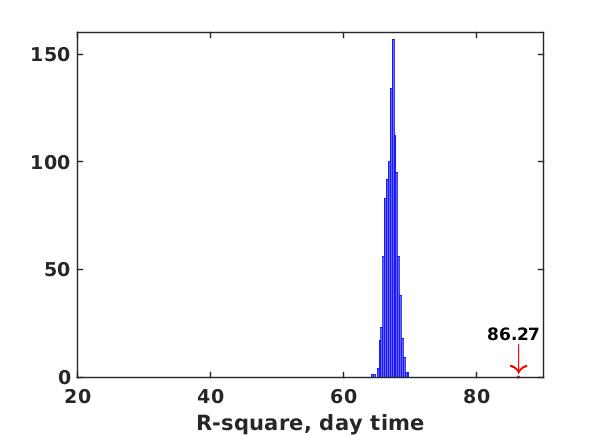}
    \includegraphics[width=7.5cm,height=2.5cm]{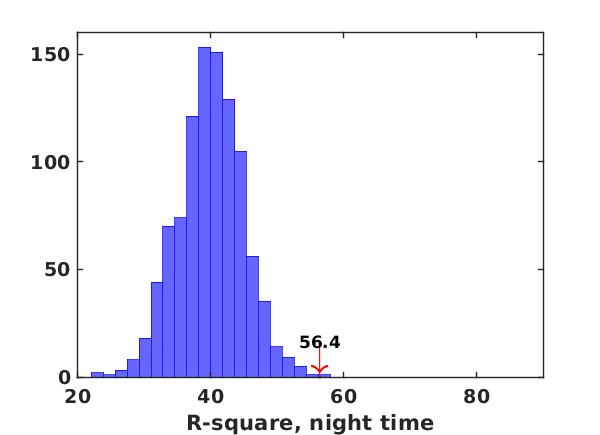}
    \caption{Permutation test to indicate the significance of $R^2$ values in all three scenarios which can be found in Table~\ref{tab:regress}.
    Each histogram is the result of randomization tests, repeated 1000 times (no. of bins = 20), where the days are shuffled before applying the regression models.}
    \label{fig:significance}
\end{figure}

After the lengthy methodology section we are now in a position to state some results. 
In order to shed light on how the price volatility could be linked to the volatility of other data of interest, we perform two pre-processing steps (discussed in detail in the previous section): 
\begin{enumerate}
\item We compute the  intra-day 2nd derivative which highlights peaks and valleys (concavity/convexity as a notion of intra-day variability). 
\item Next, we approximate these images using an SVD expansion up to rank 7. 
\end{enumerate}
We call the resulting data $C_p, C_l, C_q, C_s$ and $C_w$ where each subscript refers to the corresponding quantity.  
Finally, we regress the price data  $C_p$ on the other data: 
\begin{equation}
      C_p = \alpha_0 +  \alpha_l \hspace{.1em} C_l +  \alpha_{q}  \hspace{.1em} C_q + \alpha_s  \hspace{.1em} C_s  + \alpha_w  \hspace{.1em} C_w 
      \label{eq:regress}
\end{equation}
\begin{table}[!t]
\renewcommand{\arraystretch}{1.3} 
    \centering
 \caption{Poor results of the initial regression model on the intra-day 2nd derivative data (before using SVD-based technique)}
 \label{tab:regress01}

  \begin{tabular}{|c c c c c c |}
    \hline
  TimeSlot & $R^2$   & $\alpha_l $& $\alpha_q$ & $\alpha_s $& $\alpha_w $ \\ 
 \hline\hline
 24h &  47.28 & 1.26 &  0.44 &  -2.98 &   -1.98 \\ 
  day time  &   53.07 &   1.34 &  0.44 &  -2.97 &   -1.83 \\ 
   night time &   15.60 &   0.91 &  0.24 & N/A &  -2.05 \\ 
 \hline
 \end{tabular}    
\end{table}
The regression was performed for three scenarios on untreated (original) and enhanced (smoothed) 2nd derivative data; using all data, using only day-time data, or using only night-time data. Table~\ref{tab:regress01} contains the results for the case where the original data, $A$, were used. 
An improvement of the results was achieved by using the rank-7 reconstructed data, $A_p$, as is seen in Table~\ref{tab:regress}. Fig.~\ref{fig:significance} highlights the significance of the obtained results of the later case in comparison with the randomized data (permutation test where, first, the days of the year are shuffled, then the regression models were applied). 
\begin{table}[!t]
\renewcommand{\arraystretch}{1.3} 
\centering
    \caption{Regression model applied on rank-7 reconstruction of the intra-day 2nd derivative of data}
    \label{tab:regress}

    \begin{tabular}{|c c c c c c |}
      \hline
         TimeSlot & $R^2$   & $\alpha_l $& $\alpha_q$ & $\alpha_s $& $\alpha_w $ \\ 
      \hline\hline

      24h &  81.84 & 1.12 & 0.59 &  -2.83 &  -3.60 \\ 
      day time  &   86.27 &  1.26 & 0.29 &  -2.36 &   -1.27 \\ 
      night time &   56.40 &  0.85 &  0.09 & N/A &   -17.59 \\ 
      \hline
    \end{tabular}
\end{table}
High values of $R^2$ for \textit{24h} and \textit{day time} scenarios is an indicator of good performance of the model and emphasizes the fact that the concavity of the prices can indeed be modeled as a function of the concavity of other attributes, such as, load, traded quantity, solar and wind feed-in.

These findings lead to a number of conclusions. 
The intra-day dynamics (concavity) of the price is least affected by the traded quantity on the day-ahead market.
Moreover, RES have higher impact on the price dynamics than the load. During the day time, solar is the dominant attribute affecting the price dynamics, almost twice as much as the wind and the load.
In a similar way during night hours, wind can affect the price dynamics more than 20 times greater than the load.
In latter case, however, the lower value of $R^2$ points out to the fact that other measures are involved in the pricing mechanism at nights. 
Therefore, other attributes need to be determined to model the night time price dynamics in a more satisfactory fashion. 
%
\section{Summary} \label{sec:conclusion}
In this paper we have focused on time series in which a strong diurnal pattern is 
superimposed on slower seasonal variations. 
We have shown that it makes sense to visualize such time 
series as images/matrices, which can be approximated using low-rank SVD-based approximations. 
Furthermore, this decomposition suggests a natural structure-preserving smoothing of the data. 
We applied this decomposition to concavity data for price, load, traded quantity, 
solar and wind feed-in on the German day-ahead market (for 2016).  This process of visualization suggested a linear regression model to unveil the impact of renewables on price. 
%
%
%
%
%
%
%
%
\ifCLASSOPTIONcaptionsoff
  \newpage
\fi

\end{document}